\documentclass[aps,prx,noshowpacs,twocolumn,superscriptaddress,nobibnotes]{revtex4-1}
\usepackage{amssymb}
\usepackage{graphicx}
\usepackage{amsmath,amsfonts}
\usepackage{hyperref}
\usepackage[utf8] {inputenc}
\usepackage{xcolor}  
\usepackage[normalem]{ulem}
\usepackage[T1]{fontenc} 
\usepackage{multirow}
\usepackage{booktabs}
\usepackage{threeparttable}
\usepackage{tabularx}

\begin{document}
	\title{Geometric renormalization of weighted networks}
	\date{\today}
	\author{Muhua Zheng}
	\affiliation{School of Physics and Electronic Engineering, Jiangsu University, Zhenjiang, Jiangsu, 212013, China}
	
	\author{Guillermo Garc\'ia-P\'erez}
	\affiliation{Algorithmiq Ltd, Kanavakatu 3 C, FI-00160 Helsinki, Finland}
	
	\author{Mari\'an Bogu\~n\'a}
	\affiliation{Departament de F\'isica de la Mat\`eria Condensada, Universitat de Barcelona, Mart\'i i Franqu\`es 1, E-08028 Barcelona, Spain}
	\affiliation{Universitat de Barcelona Institute of Complex Systems (UBICS), Universitat de Barcelona, Barcelona, Spain}
	
	\author{M. {\'A}ngeles Serrano}
	\email[]{marian.serrano@ub.edu}
	\affiliation{Departament de F\'isica de la Mat\`eria Condensada, Universitat de Barcelona, Mart\'i i Franqu\`es 1, E-08028 Barcelona, Spain}
	\affiliation{Universitat de Barcelona Institute of Complex Systems (UBICS), Universitat de Barcelona, Barcelona, Spain}
	\affiliation{ICREA, Passeig Llu\'is Companys 23, E-08010 Barcelona, Spain}
	
	\begin{abstract}
		The geometric renormalization technique for complex networks has successfully revealed the multiscale self-similarity of real network topologies and can be applied to generate replicas at different length scales. In this letter, we extend the geometric renormalization framework to weighted networks, where the intensities of the interactions play a crucial role in their structural organization and function. Our findings demonstrate that weights in real networks exhibit multiscale self-similarity under a renormalization protocol that selects the connections with the maximum weight across increasingly longer length scales. We present a theory that elucidates this symmetry, and that sustains the selection of the maximum weight as a meaningful procedure. Based on our results, scaled-down replicas of weighted networks can be straightforwardly derived, facilitating the investigation of various size-dependent phenomena in downstream applications.
	\end{abstract}
	\maketitle
	\let\oldaddcontentsline\addcontentsline
	\renewcommand{\addcontentsline}[3]{}
	
	Renormalization of real networks~\cite{Garcia2018,Zheng2019a,garuccio2020multiscale,Villegas2022} can be performed on a geometric framework~\cite{Garcia2018} by virtue of the discovery that their structure is underlain by a latent hyperbolic geometry~\cite{boguna2020network,serrano_boguna_2022}. Distances between nodes in this space determine the likelihood of connections via a universal law that operates at all scales and encodes simultaneously short- and long-range connections. This geometric principle has been able to explain many features of real networks, including the small-world property, scale-free degree distributions, and high levels of clustering, as well as fundamental mechanisms such as preferential attachment in growing networks~\cite{Papadopoulos2012}, and the emergence of communities~\cite{Garcia2018aa,Zuev2015aa}. It has also led to embedding techniques that produce geometric representations of complex network from their topologies~\cite{Boguna2010,Papadopoulos:2015ub,muscoloni2017machine,blasius2018efficientembedding,Garcia2019}. 
	
	Weights in real complex networks~\cite{Barrat:2004b,newman2004analysis,Serrano:2006fu,mastrandrea2014enhanced,menichetti2014weighted} are also amenable to modeling within the hyperbolic network geometry paradigm. More specifically, the weighted geometric soft configuration model (W$\mathbb{S}^D$)~\cite{allard2017geometric} captures the non-trivial coupling between network topology and weights, allowing for accurate reproduction of both the unweighted and the weighted structure of real networks. However, the geometric renormalization (GR) method only applies to unweighted networks. By applying coarse graining and rescaling steps to unfold an unweighted network map into a sequence of scaled-down layers over progressively longer length scales, GR revealed multiscale self-similarity to be a ubiquitous symmetry in real networks~\cite{Garcia2018}. This raises the question whether GR can be generalized to weighted networks as well and whether self-similarity would be preserved in that case. 
	
	Adding to GR, the geometric renormalizaton of weights (GRW) should produce the multiscale unfolding of a network into a shell of weighted scaled-down layers that preserve the weighted structure of the network in the flow. Here, we propose a theory for the renormalization of weighted networks that supports the selection of the maximum, or supreme, as an effective approximation to allocate weights in the renormalized layers of real networks. Our theory is sustained by the renormalizability of the W$\mathbb{S}^D$ model, which entails that the GRW transformation should be a rescaled $p$-norm on the set of weights to be renormalized.
	Alternatively, the GR technique was recently extended to weighted networks using an ad hoc approach that treats weights as currents or resistances in a parallel circuit---renormalizing by the sum of the weights or by the inverse of the sum of their inverses, respectively~\cite{9761989}. The two methods are recovered as particular limits of our theory.

	
	\begin{figure*}[t]
		\centering
		\includegraphics[width=1\linewidth]{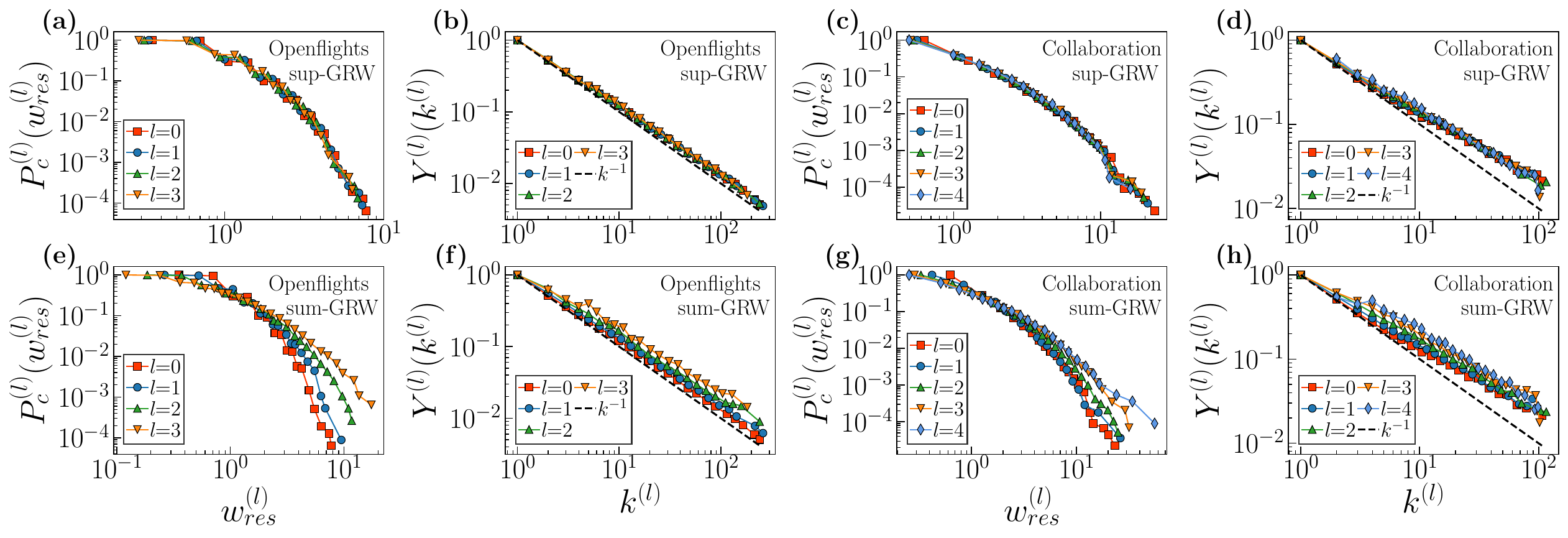}
		\caption{{\bf Self-similarity of real weighted networks along GRW flows.} The first row shows the sup-GRW flow of the pdf of weights and their disparity in nodes in Openflights (a)-(b) and Collaboration (c)-(d). The same in the second row for the sum-GRW flow. The number of layers in each shell is determined by the original network size, and $r=2$ in all cases. } 
		\label{fig:properties}
	\end{figure*}
	
	To begin with, we provide evidence that self-similarity is a pervasive symmetry not only in the multiscale organization of real network topologies but also in the multiscale ulfonding of their weights. 
	To that end, we implement a GRW transformation, which requires the preliminary application of the GR technique to unweighted networks~\cite{Garcia2018}. 
	
	The GR technique operates on the geometric embedding of a network, as described in previous works~\cite{Boguna2010,Garcia2019}, obtained by maximizing the likelihood that the network topology is generated by the geometric soft configuration model $\mathbb{S}^D$~\cite{Serrano2008}. In this model, nodes are assigned coordinates representing popularity and similarity dimensions, and distances between them determine the probability of connection $p_{ij}=1/(1+\chi_{ij}^\beta)$, where $\chi_{ij}=
	d_{ij}/(\mu\kappa_i\kappa_j)^{1/D}$. Parameter $\mu$ controls the average degree, and $\beta>D$ controls the level of clustering and quantifies the level of coupling between the network topology and the geometry. The hidden degree $\kappa_i$ of node $i \in [1,N]$---equivalent to a radial coordinate in the hyperbolic plane in the purely geometric formulation of the model, named $\mathbb{H}^{D+1}$~\cite{Krioukov2009}---measures the popularity of the node, with higher values indicating a greater likelihood of connecting to other nodes. In $D=1$, the similarity subspace is represented as a circle of radius $R = N / (2 \pi)$ with unit density. Each node $i$ is assigned an angular coordinate $\theta_i$ in the circle, and angular distances $d_{ij} = R \Delta\theta_{ij}$ between pairs of nodes account for factors other than degrees that influence the tendency to form connections. Nodes closer in the similarity subspace have a higher likelihood of being connected. Hyperbolic embeddings of unweighted networks can be obtained using the Mercator mapping tool~\cite{Garcia2019}, which employs statistical inference techniques to identify the hidden degrees and angular coordinates while adjusting parameters $\beta$ and $\mu$ accordingly.
	
	Once the geometric map of a real network is generated, GR divides the similarity circle into non-overlapping blocks of consecutive nodes of size $r$. These blocks are then coarse grained forming supernodes in a new layer. Each supernode is positioned within the angular region defined by the corresponding block, preserving the order of nodes. Any links between nodes in one supernode and nodes in another are renormalized into a single link connecting the two supernodes. This way, GR eliminates short-range couplings and produces a new network topology that is self-similar to the original except for the average degree, which
	increases in the renormalization flow~\cite{Garcia2018}. 
	
	The GRW technique involves assigning intensities to the links in the new layer based on the weights in the original layer, following a specific prescription. This transformation can be iterated starting from the original network at layer $l=0$, with the iteration bounded to approximately $l_{max} \propto \log(N)$ steps due to the finite size of real networks. As a result, a sequence of self-similar network layers $l$---each $r$ times smaller than the original one---is produced forming a multiscale weighted shell of the original network. The process is visually depicted in Fig.~S1 of the Supplemental Material (SM). The crux of GRW lies in how the weights are renormalized
	to ensure that their characteristics, such as global and local weight distributions and the relationship between strength and degree, are preserved throughout the renormalization flow.
	
	An effective and simple prescription, referred to as sup-GRW, is to define the weight of the link between two supernodes as the maximum, or supremum, of the weights in the existing links between their constituent nodes in the original layer. We applied the sup-GRW technique to $12$ different real weighted networks from different domains including biology, transportation, knowledge, and social systems. The networks were processed using blocks of size $r = 2$. Additional details can be found in the SM.
	
	The behavior of the weights in the renormalization flow of two of the networks are shown in Fig.~\ref{fig:properties}(a)-(d), while Figs.~S2-S5 present the corresponding results for the remaining networks. The relations strength-degree are shown in Fig. S5. The probability density functions (pdf) of weights and strengths in the different layers collapse once rescaled by the average weight and average strength, respectively, in the corresponding layer. Furthermore, the power-law relations between strength and degree also overlap once the degrees are rescaled by the average degree of the layer, as demonstrated in Figs.~S4 and S5. To quantify the local heterogeneity of the weights, we measured their disparity around nodes as a function of the degree, as described in the Methods section of the SM. The results show, again, statistical invariance across layers.
	
	Notice that, by construction, the average weight and the average strength in the sup-GRW layers grows with $l$. While this behavior does not provide fundamental information for characterizing the description of the weighted structure of the network, it may still be interesting to understand how $\langle w \rangle$ and $\langle s \rangle$ depend on the scale of observation $l$. This is particularly relevant considering that weights in real networks are often expressed in real-world units. The corresponding results are presented in Figs.~S6 and S7. Furthermore, the sup-GRW transformation exhibits the semigroup structure with respect to the composition, similar to the behavior observed in GR for unweighted networks. This means that a certain number of iterations with a given coarse graining factor are equivalent to a single transformation with a higher coarse graining factor. The findings shown in Fig.~S8 provide support for this claim.
	
	We also tested an alternative prescription, referred as sum-GRW, where weights in the new layer are assigned by summing the weights of existing links between the nodes in supernodes, following the prescription described in Ref.~\cite{9761989}. While this strategy proves effective for many real networks, there are certain cases in which self-similarity is not maintained in the renormalization flow. When sum-GRW is applied, the global distribution of weights, the local heterogeneity of weights in nodes, and the relation between strength and degree become increasingly heterogeneous compared to the original graph. This is observed in the Openflights and the scientific collaboration network, as illustrated in Fig.~\ref{fig:properties}(e)-(h), and in Figs.~S9 and S10 for the remaining  networks.

	The reported results are supported by a theoretical framework that clarifies the conditions under which each of the two weight assignment prescriptions, selecting the
	supremum of weights between supernodes or their sum, yields good performance. Our theory is based on the W$\mathbb{S}^D$ model~\cite{allard2017geometric}, that uses the $\mathbb{S}^D$ model to mimic the topology of real networks. In the W$\mathbb{S}^D$ model, weights are assigned to connections between two connected nodes $i$ and $j$ as follows:
	\begin{equation}\label{eq:wij}
		\omega_{ij} = \epsilon_{ij} \frac{\nu \sigma_i \sigma_j}{\left( \kappa_i \kappa_j \right)^{1 - \alpha/D} d_{ij}^{\alpha}}.
	\end{equation}
	Similar to $\bar{k}_i \propto \kappa_i$ in the $\mathbb{S}^D$ model, the W$\mathbb{S}^D$ model ensures that the expected strength of node $i$, $\bar{s}_i$, is proportional to the hidden strength $\sigma_i$, $\bar{s}_i \propto \sigma_i$. When $\alpha=0$, the weights are independent of the underlying geometry and primarily influenced by node degrees, while $\alpha=D$ implies that weights are maximally coupled to the underlying metric space with no direct contribution of the degrees. Finally, $\epsilon_{ij}$ is a random variable with mean equal to one and the variance of which regulates the level of noise in the network. In the subsequent analysis, we assume the noiseless version of the model to simplify analytical calculations, which means $\epsilon_{ij}=1 \; \forall (i,j)$.
	
	To control the correlation between strength and degree and, consequently, adjust the strength distribution, we assume a deterministic relation between hidden variables $\sigma$ and $\kappa$ of the form $\sigma = a \kappa^\eta$, yielding $s(k) \sim ak^{\eta}$ as observed in real complex
	networks. Working under this assumption, a valid GRW transformation should preserve the relation between strength and degree, and in particular the exponent $\eta$, meaning that the renormalized hidden degree and strength should satisfy $\sigma' = a' (\kappa')^{\eta}$ (to simplify notation, we have used prima to denote quantities in the renormalized layer). Using Eq.~\eqref{eq:wij} and the GR equations for the topological model~\cite{Garcia2018}, this requirement leads to the following expression for the renormalized weights
	\begin{equation}\label{eq:transformation}
		\omega'_{ij} = C\left[ \sum\limits_{e=1}^{r^2} \left( w_{mn} \right)_e^{\phi} \right]^{1/\phi} ,
	\end{equation}
	where the sum runs over the links between nodes within supernodes $i$ and $j$, derivation in SM. Parameter $\phi \equiv \frac{\beta}{D(\eta - 1) + \alpha}$ depends on both the weighted and unweighted structure of the network, and $C=\nu'/\nu \left( a'/a \right)^2 r^{\alpha/D}$. In practice, however, we rescale weights by the average weight in each layer, rendering the constant $C$ irrelevant.
	
	According to the weighted model, for a network with a specific value of $\phi$, the GRW transformation of weights Eq.~(\ref{eq:transformation}), denoted as $\phi$-GRW, preserves the exponent $\eta$ that characterizes the relation between strength and degree. At the same time, since the distribution of hidden degrees is assumed to be preserved by GR, the distribution of hidden strengths and the distribution of weights are also preserved. This is valid as long as $\beta > (\gamma - 1) / 2$. Otherwise, the power-law distribution of hidden degrees looses its self-similarity in the unweighted  renormalization  flow and this breaks the self-similarity of weights. Also, note that the $\phi$-GRW transformation has semigroup structure with respect to the composition, regardless of the value of $\phi$.
	
	
	We validated the self-similarity of the $\phi-$GRW transformation in the real and synthetic networks, Figs.~S11-S12 and Figs.~S14-S17, respectively, including its semigroup property. In all cases, the self-similar behavior of the distribution of weights and strengths, and the power-law relation between strength and degrees in the renormalization flow is clear across length scales, which validates our analytic calculations.

	Notice that the transformation in Eq.~(\ref{eq:transformation}) is a $\phi$-norm, which is a generalization of the Euclidean norm. As $\phi$ increases, the $\phi$-norm becomes progressively dominated by the supremum of the terms $w_{mn}$ in Eq.~(\ref{eq:transformation}) . In fact, the sup-GRW prescription is recovered in the limit $\phi=\infty$ of $\phi$-GRW. In addition, renormalizing by the sum is equivalent to setting $\phi=1$, and the renormalization of weights by the inverse of the sum of inverse values corresponds to $\phi=-1$.
	
	\begin{figure}[!t]
		\includegraphics[width=1\linewidth]{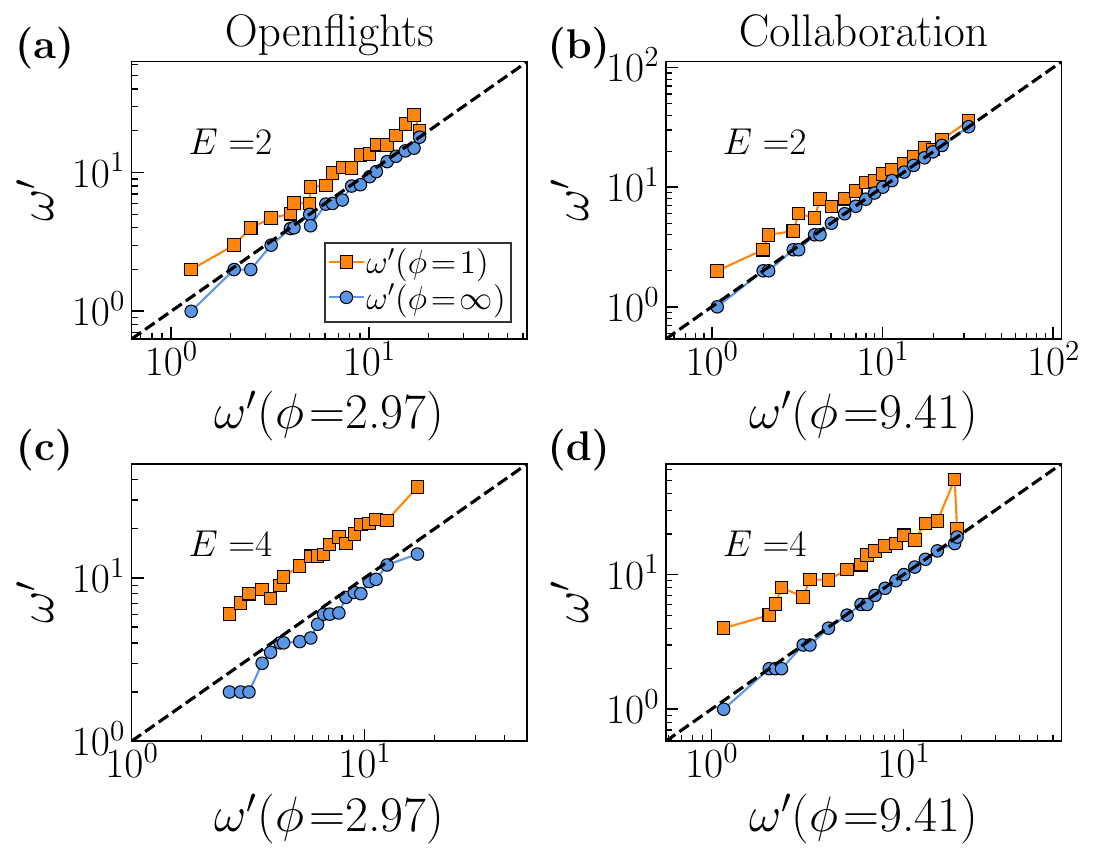}  
		\caption{ {\bf  Asymptotics of the $\phi$-norm.} We used the weights $\{\omega_{mn}\}$ in the Openflights and Collaboration networks, and performed an iteration of $\phi-$GRW to calculate the renormalized weight $\omega'$ with Eq.~(\ref{eq:transformation}). Note that when $r=2$, the number of links $E$ between the nodes in two supernodes could be $1$, $2$, $3$ or $4$. So we displayed the renormalized weight $\omega'(\phi=1)$ and $\omega'(\phi=\infty)$ versus $\omega'(\phi^*)$ for different $E$, where $\phi^*$ is the inferred value with $\phi^*=\beta/(\eta-1+\alpha)$. Sup-GRW corresponds to the case $\phi=\infty$ while sum-GRW to $\phi=1$.}
		\label{fig:asymptotics}
	\end{figure}
	
	To clarify the efficacy of approximating $\phi-$GRW as sup-GRW, we checked the asymptotic behavior of the $\phi$-norm as a function of the number of elements $E$ in the set of coarse-grainable weights and of the level of heterogeneity in the weights, see section VI in SM for more details. Figure \ref{fig:asymptotics} shows the result of applying the supremum and the sum prescriptions as compared with renormalizing weights using $\phi$-GRW in two of the real networks analyzed in this letter, see Figs.~S20 and S21 for the rest. In synthetic networks, we simulated weights using a distribution $p(\omega_{mn})\sim \omega_{mn} ^{-\delta}$, where $\delta$ allowed us to tune the level of heterogeneity, and produced sets of weights that were renormalized using Eq.~(\ref{eq:transformation}) with $C=1$ and different values of $\phi$. We also renormalized the same sets using the alternative sum and supremum prescriptions, the results are shown in Figs.~S18 and S19. 
	
	In heterogeneous networks with a markedly scale-free character of the weight distribution, very small deviation from the supremum are observed and this occurs primarily for very low values of $\phi$ and low-weight values. As the number of elements $E$ increases and the degree distribution becomes more homogeneous, these deviations progressively become larger. As expected, higher values of $\phi$ reduce the discrepancy between the $\phi$-norm and the supremum estimator. Nevertheless, across
	a wide range of parameter values, which encompass those for realistic networks, there is generally a good agreement between the $\phi$-norm and the selection of the supremum, with any existing deviations being quite minor. While for some empirical weight distributions, sup-GRW and sum-GRW yield the same renormalized weights, e.g., the JCN in Figs.~S20 and S21, it is important to note that, in general, the relation between hidden strength and hidden degree is not preserved under sum-GRW. See Methods section in SM for more details.

	\begin{figure}[!b]
		\includegraphics[width=1\linewidth]{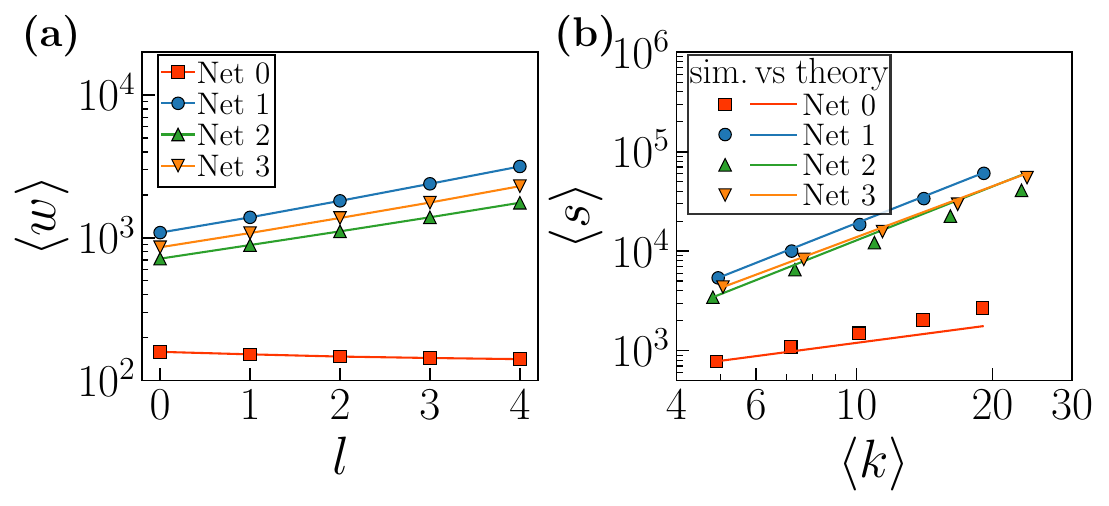}  
		\caption{{\bf Analytic approximation vs simulations.} (a) Unrescaled average weight for different layers $l$. (b) Average strength, $\langle s\rangle$ as a function of average degree $\langle k\rangle$, in which symbols are the simulated results and lines indicate the corresponding theoretical analysis from Eq.\eqref{eq:s_vs_k}. 
			The synthetic networks are generated with $\gamma=2.7$, $\beta=1.5$, $\alpha=0.4$ and $\eta=1.2$ 
			for Net $0$; $\gamma=2.7$, $\beta=1.5$, $\alpha=0.4$ and $\eta=1.8$ 
			for Net $1$; $\gamma=2.2$, $\beta=2.0$, $\alpha=0.4$ and $\eta=1.5$ 
			for Net $2$; $\gamma=2.2$, $\beta=2.0$, $\alpha=0.6$ and $\eta=1.5$ 
			for Net $3$. In all simulations $N=64000$, $\langle k \rangle=5$, $a=100$, and $\langle \epsilon^2 \rangle=1.0$.} 
		\label{fig:c_k_w_s}
	\end{figure}
	The preservation of the relation $\sigma = a \kappa^{\eta}$ allows us to approximate analytically the flow of the average strength from the flow of the average degree. In GR, the average degree changes from layer to layer approximately as $\langle k \rangle^{(l+1)}=r^{\xi} \langle k \rangle^{(l)}$, with a scaling factor $\xi$ depending on the connectivity structure of the original network~\cite{Garcia2018}. Combining this with Eq.~(\ref{eq:transformation}) and imposing that the rescaling constant of weights $C$ does not change in the flow, we obtain 
	\begin{equation}
		\langle \sigma' \rangle=  \langle \sigma \rangle r^{\psi}, \mbox{      }\psi=\left(\frac{\alpha}{D}+2 \eta - 1\right)\xi - \frac{\alpha}{D},
		\label{eq:renors}
	\end{equation}
	which, due to the proportionality between observed and hidden strength, implies that the flow of the average observed strength follows the same scaling.
	Therefore, in $D=1$, the strength increases with a scaling factor that depends on the exponent $\eta$, on the coupling $\alpha$ between topology and geometry, and on the scaling factor $\xi$ for the flow of the average degree, see Methods in SM for details. This leads to an analytic approximation for the growth of the average strength as a function of the average degree
	\begin{equation} 
		\langle s \rangle^{(l)}=\langle s \rangle^{(0)} \left( \frac{\langle k \rangle^{(l)}}{\langle s \rangle^{(0)}}\right) ^{\frac{\psi}{\xi}},  
		\label{eq:s_vs_k}
	\end{equation}	
	which agrees with the measurements in synthetic networks where the average weight may increase, stay flat, or decrease in the flow as shown in Fig.~\ref{fig:c_k_w_s}. 
	
	
	All together, our results suggest that sup-GRW is a good approximation for real networks and offers certain advantages over $\phi$-GRW. One advantage is that it avoids the need to estimate parameters that capture the coupling between the weighted structure of the network and the underlying geometry, which can be challenging in practice. Sup-GRW is equivalent to setting $\phi=\infty$ and, due to the nature of the transformation, it is effectively reached for relatively low values of $\phi$. In addition, renormalizing by the sum is equivalent to setting $\phi=1$, which in general does not preserve the exponent $\eta$ of the relation between $\sigma$ and $\kappa$, see the Methods section in SM for analytical calculations. 
	
	Beyond theoretical considerations, the practical application of GRW extends to the generation of scaled-down replicas of weighted networks. These replicas can serve as
	valuable testbeds for evaluating the scalability of computationally intensive protocols or studying processes where the size of a real network plays a role.
	The generation of a scaled-down replica involves obtaining a reduced version of the topology, as described in Ref.~\cite{Garcia2018}, and subsequently rescaling the weights in the renormalized network layer to mathc the level of the original network. The detailed procedure can be found in the SM, and the results for the scaled-down replicas of real weighted networks are presented in Figs. S25-S28.
	
	
	In summary, the extension of the geometric renormalization framework to weighted networks demonstrates that multiscale self-similarity characterizes not only the
	topology but also the weighted structure of real networks, provided the appropriate renormalization scheme is applied. Moreover, the weights in these networks result from processes that determine the intensities of interactions, and our findings suggest that these processes follow the same underlying principles across different length scales. Notably, the transformation implied by the theory is closely approximated by using the maximum weight prescription, a highly effective approach that can be readily applied to real networks despite the presence of significant noise affecting their weights. This observation justifies our confidence that noise will not fundamentally alter the qualitative results reported in this study.
	
	The present work represents a significant step towards establishing a comprehensive framework for the renormalization of network structure and opens up possibilitis for renormalizing dynamical processes on real networks. In future research, it will be essential to incorporate not only the topology of connections and their weights but also their directionality, which is crucial in many real-world processes.

	

	%
	We thank Elisenda Ortiz for helpful discussions. M.Z. acknowledges support from National Natural
	Science Foundation of China (Grants No.~12005079), the Natural Science Foundation of Jiangsu Province (Grant No.~BK20220511), 
	the funding for Scientific Research Startup of Jiangsu University (Grant No.~4111710001), and Jiangsu Specially-Appointed Professor Program. M.~A.~S and M.~B. acknowledge support from the Agencia Estatal de Investigaci\'on project number PID2019-106290GB-C22 funded by MCIN/AEI/10.13039/501100011033; Generalitat de Catalunya grant number 2021SGR00856. M.~B. acknowledges support from the ICREA Academia award, funded by the Generalitat de Catalunya.


%

\onecolumngrid
\appendix
\renewcommand\appendix{\setcounter{secnumdepth}{0}}
\newpage
\clearpage

\setcounter{page}{0}
\pagenumbering{arabic}

\clearpage
\begin{minipage}[h]{\textwidth}
	\begin{center}
		\large{\textbf{Supplemental Material for\\ Geometric renormalization of weighted networks}}\\
		\vspace{0.5cm}
		Muhua Zheng$^{1}$, Guillermo Garc\'ia-P\'erez$^{2}$, Mari\'an Bogu\~n\'a$^{3,4}$ \& M. \'Angeles Serrano$^{3,4,5*}$\\ 
		\vspace{1cm}		
		\small{$^{1}$\textit{ School of Physics and Electronic Engineering, Jiangsu University, Zhenjiang, Jiangsu, 212013, China}}\\		
		\small{$^{2}$\textit{Algorithmiq Ltd, Kanavakatu 3 C, FI-00160 Helsinki, Finland}}\\		
		\small{$^{3}$\textit{Departament de F{\'\i}sica de la Mat\`eria Condensada,\\ Universitat de Barcelona, Mart\'{\i} i Franqu\`es 1, 08028 Barcelona, Spain}}\\
		\small{$^{4}$\textit{Universitat de Barcelona Institute of Complex Systems (UBICS), \\Universitat de Barcelona, Barcelona, Spain}}\\
		\small{$^{5}$\textit{ICREA, Pg. Llu\'is Companys 23, E-08010 Barcelona, Spain}}\\			
		\small{\textit{*Correspondence and requests for materials should be addressed to M.A.S. (marian.serrano@ub.edu)}}
		\\

	\end{center}
	
\end{minipage}

\vspace{2cm}



\tableofcontents
\let\addcontentsline\oldaddcontentsline

\newpage
\renewcommand\thefigure{S\arabic{figure}}
\renewcommand\thetable{S\arabic{table}}
\setcounter{figure}{0}    
\setcounter{table}{0}

\renewcommand{\appendixname}{}

\section{Illustration of the geometric renormalization transformation method for weighted networks}
\begin{figure}[!h]
	\centering
	\includegraphics[width=0.6\linewidth]{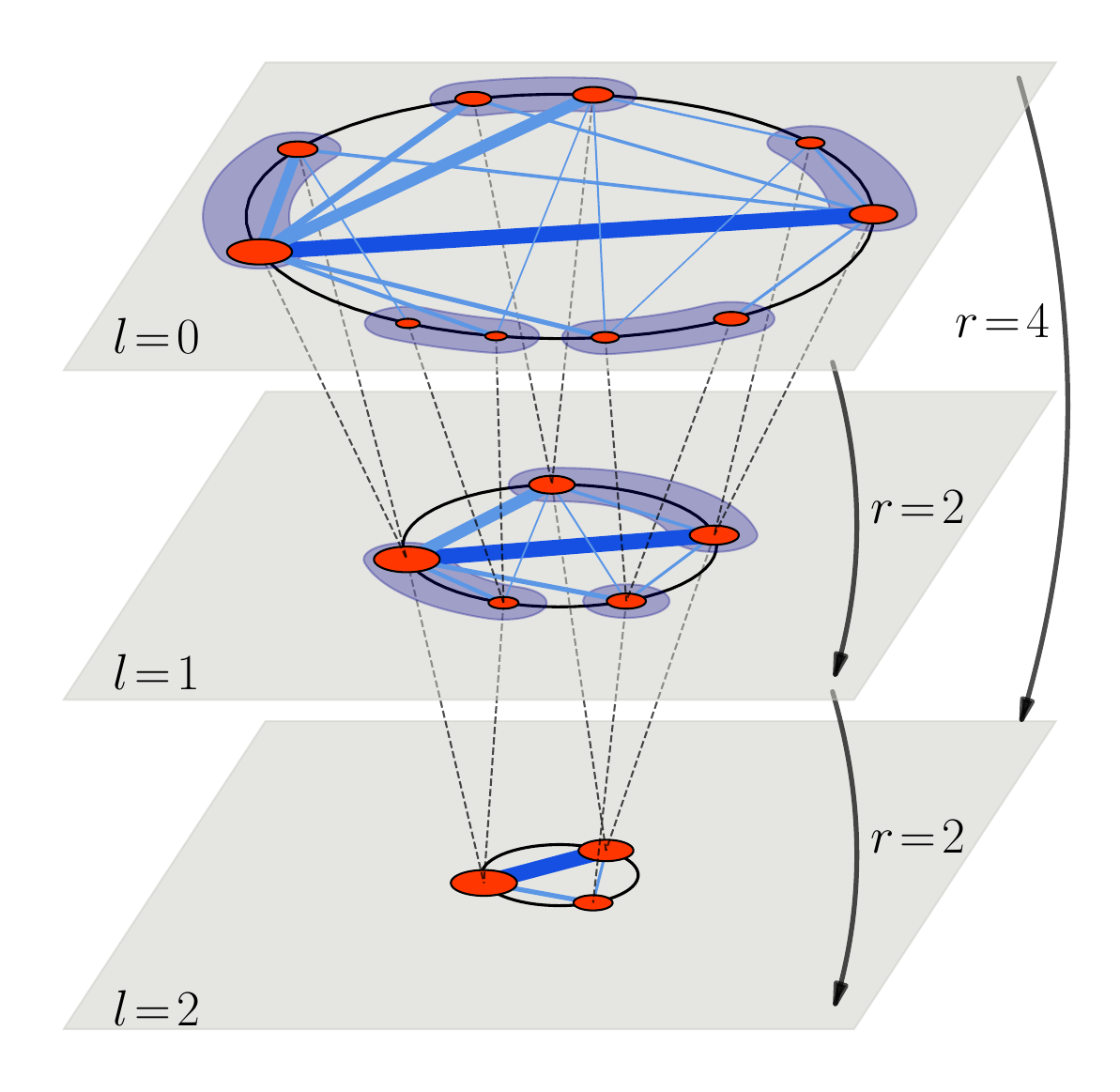}
	\caption{ {\bf Geometric renormalization transformation for weighted networks.} Each layer is obtained after a GRW step with resolution $r$ starting from the original network in $l=0$. Each node $i$ in red is placed at an angular position on the similarity circle and has a size proportional to the logarithm of its hidden degree. Straight solid lines represent the links in each layer with weights denoted by their thickness. Coarse-graining blocks correspond to the blue shadowed areas, and dashed lines connect nodes to their supernodes in layer $l+1$. Two supernodes in layer $l+1$ are connected if and only if some node of one supernode in layer $l$ is connected to some node of the other, with the supremum among the weights of links between the constituent nodes as the weight of the new connection (dark blue links give an example). The GRW transformation has semigroup structure with respect to the composition. In the figure, the transformation with $r=4$ goes from $l=0$ to $l=2$ in a single step. 
	}
	\label{fig:sketch}
\end{figure}

%
\section{Methods}

\subsection{Description of empirical data sets}
\label{A}

\begin{itemize}
	\item\textbf{Cargo ships}.
	The international network of global cargo ship movements consists of the
	number of shipping journeys between pairs of major commercial ports in the world
	in 2007~\cite{kaluza2010}.
	
	\item\textbf{E. coli}.
	Weights in the metabolic network of the bacteria {\it E. coli} K-12 MG1655 consist
	of the number of different metabolic reactions in which two metabolites
	participate~\cite{serrano2012,orth2011}.
	
	\item\textbf{US commute}.
	The commuting network reflects the daily flow of commuters between counties
	in the United States in 2000~\cite{grady2012}.
	
	\item\textbf{Facebook like Social Network(Facebook)}.
	The Facebook-like Social Network originate from an online community for students at University of California, Irvine, in the period between April to October 2004~\cite{panzarasa2009,opsahl2009}. In
	this network, the nodes are students and ties are established when online messages are exchanged between the students. The weight of a directed tie is defined as the number of messages sent from one student to another. We discard the directions for any link and preserve the weight $\omega_{ij}$ with the sum of bidirectional messages, i.e., $\omega_{ij}=\omega_{i \to j}+\omega_{j\to i}$. Notice that we only consider the giant connected component of the undirected and weighted networks in this paper.

	\item\textbf{Collaboration}. 
	This is the co-authorship network of based on preprints posted to Condensed Matter section of arXiv E-Print Archive between 1995 and 1999~\cite{newman2001}. Authors are identified with nodes, and an edge exists between two scientists if they have coauthored at least one paper. The weights are the sum of joint papers. Notice that we only consider the giant connected component of the undirected and weighted networks in this paper.
	
	\item\textbf{Openflights}.
	Network of flights among all commercial airports in the world, in 2010, 
	derived from the Openflights.org database~\cite{Opsahl2011}. Nodes represent the airports. The weights in this network refer to the number of routes between two airports. We discard the directions for any link and preserve the weight $\omega_{ij}$ with the sum of bidirectional weights, i.e., $\omega_{ij}=\omega_{i \to j}+\omega_{j\to i}$. Notice that we only consider the giant connected component of the undirected and weighted networks in this paper.
	
	\item\textbf{Journal Citation Network (JCN)}. The citation networks from 1900 to 2013 were reconstructed from data on citations between scientific articles extracted from the Thomson Reuters Citation Index~\cite{Hric2018}. A node corresponds to a journal with publications in the given time period. An edge is connected from journal $i$ to journal $j$ if an article in journal $i$ cites an article in journal $j$, and the weight of this link is taken to be the number of such citations. In this work, we use undirected and weighted networks generated from 3 different time windows, 2008-2013, 1985-1990 and 1965-1975. The data are obtained from Ref.~\cite{zheng2021}.
	
	\item\textbf{New Zealand Collaboration Network (NZCN)}. This is a network of scientific collaborations among institutions in New Zealand. Nodes are
	institutions (universities, organizations, etc.) and edges represent collaborations between them. In particular, two nodes $i$, $j$ are connected if Scopus lists at least one publication with authors at institutions $i$ and $j$, in the period 2010-2015. The weights of edges record the number of such collaborations. The data are obtained from Ref.~\cite{Aref2018analysing}. Notice that we only consider the giant connected component of the undirected and weighted networks in this paper.
	
	\item\textbf{Poppy and foxglove hypocotyl cellular interaction networks}. 
	These networks capture global cellular connectivity within the hypocotyl (embryonic stem) of poppy and foxglove. Nodes represent cells and edges are their physical associations in 3D space. Edges are weighted by the size of shared intercellular interfaces, and nodes annotated with cell type. The data are obtained from Ref.~\cite{Jackson2017}.  
\end{itemize}

Network statistics can be found in Table S1. 
\renewcommand{\arraystretch}{1} 
\newcommand*{\thd}[1]{\multicolumn{1}{l}{#1}}
\begin{table*}[!h] 
	\begin{ruledtabular}
		\centering		
		\caption{Overview of the considered real-world networks. Columns are: the name  of each network (Name), the number of nodes ($N$), the average degree ($\langle k \rangle$), the average local clustering coefficient ($\langle c \rangle$), the hyperbolic embedding parameter $\beta$ and $\mu$, fitting exponent ($\gamma$) in degree distribution, fitting parameters ($a$ and  $\eta$) in strength-degree relations, the trade-off between
			the contribution of degrees and geometry to weights ($\alpha$), the noise ($\langle\epsilon^2\rangle$), parameter $\phi=\beta/(\eta-1+\alpha)$, and the references about the data sources (Ref.). }
		\label{tab:networkproperties}		
		\begin{tabular}{*{14}{l}}
			\thd{No.}	&\thd{Name}  &\thd{$N$} &\thd{$\langle k\rangle$}  &\thd{$\langle c\rangle$} &\thd{$\beta$} &\thd{$\mu$} &\thd{$\gamma$} &\thd{$a$} &\thd{$\eta$} &\thd{$\alpha$} &\thd{$\langle\epsilon^2\rangle$} &\thd{$\phi$} & \thd{Ref.}
			\\
			\hline	
			0      &Cargo ships      &821    &10.58      &0.51       &1.88       &0.028      &3.3        &86.10      &1.04       &0.66       &1.50       &2.68      &~\cite{kaluza2010}\\ 
			1      &E. coli          &1100   &6.61       &0.49       &1.98       &0.048      &2.7        &1.09       &1.10       &0.45       &1.30       &3.65      &~\cite{serrano2012,orth2011}\\ 
			2      &US commute       &3025   &4.36       &0.40       &2.08       &0.076      &5.5        &696.78     &2.06       &0.57       &1.50       &1.28     &~\cite{grady2012} \\ 
			3      &JCN(2008-2013)   &21460  &49.79      &0.59       &1.70       &0.005      &3.4        &7.35       &1.41       &0.62       &1.30       &1.65      &~\cite{Hric2018,zheng2021}\\ 
			4      &JCN(1985-1990)   &7379   &28.46      &0.48       &1.63       &0.009      &2.7        &9.35       &1.48       &0.46       &1.50       &1.74      &~\cite{Hric2018,zheng2021}\\ 
			5      &JCN(1965-1975)   &4168   &26.84      &0.61       &1.93       &0.011      &3.0        &17.52      &1.49       &0.18       &1.70       &2.90      &~\cite{Hric2018,zheng2021}\\ 
			6      &Facebook &1893   &14.62      &0.14       &1.01       &0.001      &2.9        &1.97       &1.20       &0.45       &1.10       &1.56      &~\cite{panzarasa2009,opsahl2009}\\ 
			7      &Collaboration    &13861  &6.44       &0.72       &5.42       &0.073      &3.8        &1.32       &1.08       &0.50       &1.10       &9.41      &~\cite{newman2001}\\ 
			8      &Openflights      &2905   &10.77      &0.59       &1.87       &0.027      &1.9        &2.16       &1.08       &0.55       &1.00       &2.97      &~\cite{Opsahl2011}\\ 
			9      &NZCN &1463   &5.80       &0.85       &14.33      &0.085      &2.3        &0.90       &1.50       &0.80       &1.80       &11.00     &~\cite{Aref2018analysing}\\ 
			10     &Poppy            &2507   &11.51      &0.43       &2.31       &0.031      &16.9       &26.23      &1.20       &0.35       &1.70       &4.20      &~\cite{Jackson2017}\\ 
			11     &Foxglove         &3005   &11.34      &0.43       &2.33       &0.032      &13.0       &14.69      &1.46       &0.39       &1.70       &2.72      &~\cite{Jackson2017}\\ 
			
		\end{tabular}
	\end{ruledtabular}
\end{table*}

\subsection{Network embedding to produce geometric network maps}
\label{B}

We embed each considered network into hyperbolic space using the algorithm introduced in Ref.~\cite{Garcia2019}, named Mercator. Mercator takes the network adjacency matrix $A_{ij}$ ($A_{ij}=A_{ji}=1$ if there is a link between nodes $i$ and
$j$, and $A_{ij}=A_{ji}=0$ otherwise) as input and then returns inferred hidden degrees, angular positions of nodes and global model parameters. More precisely, the hyperbolic maps were inferred by finding the hidden degree and angular position of each node, $\{\kappa_i\}$ and $\{\theta_i\}$, that maximize the likelihood $\mathcal{L}$ that the structure of the network was generated by the $\mathbb{S}^1$ model, where
\begin{align}
	\mathcal{L} = \prod_{i<j} \left[ p_{ij} \right]^{A_{ij}} \left[ 1 - p_{ij} \right]^{1 - A_{ij}} \ ,
\end{align}
and $p_{ij}=1/(1+\chi_{ij}^\beta)$ is the connected probability.

\subsection{The definition of disparity}
\textbf{The disparity of nodes.} The disparity quantifies the local heterogeneity of the weights attached to a given node $i$ and is defined as 
\begin{equation}\label{eq:Yi}
	\begin{aligned}
		Y(k_i)=\sum_j\left( \frac{\omega_{ij}}{\sum_j \omega_{ij}}\right) ^2
	\end{aligned}
\end{equation}
where $\omega_{ij}$ is the weight of the link between node $i$ and its neighbor $j$. From this definition, we see that the disparity scales as $Y \sim k_i^{-1}$, whenever the weights are roughly homogeneously distributed among the
links. Conversely, whenever the disparity decreases slower than $k_i^{-1}$ implies that
weights are heterogeneous and that the large strength of a node is due to a handful
of links with large weights.


\subsection{Theoretical derivation of the renormalized weights}
\label{C}

Under GR, the hidden variables of supernodes in the resulting layer, $\kappa^\prime$ and $\theta^\prime$, are calculated as a function of the hidden variables of the constituent nodes as
\begin{eqnarray} \label{eq:RG_kappa_theta}
	\kappa'=\left[  \sum_{j=1}^{r}(\kappa_j)^\beta \right] ^{1/\beta} \text{and  \hspace{0.1cm} }
	\theta'=\left[  \frac{\sum_{j=1}^{r}(\theta_j\kappa_j)^\beta }{\sum_{j=1}^{r}(\kappa_j)^\beta}\right] ^{1/\beta}. \ 
\end{eqnarray}

The expressions above and Eq. (1) in main text altogether imply that the renormalized weight should be
\begin{equation}
	\begin{aligned}
		\omega'_{ij} &= \frac{\nu' \sigma'_i \sigma'_j}{\left( \kappa'_i \kappa'_j \right)^{1 - \alpha'/D} d_{ij}^{'\alpha'}} = \nu' (a')^2 d_{ij}^{'-\alpha'} \left( \kappa'_i \kappa'_j \right)^{\eta - 1 + \alpha'/D} \\
		&= \epsilon'_{ij} \nu' (a')^2 d_{ij}^{'-\alpha'} \left[ \left( \kappa'_i \kappa'_j \right)^{\beta/D} \right]^{\frac{D(\eta - 1) + \alpha'}{\beta}} \\
		&= \nu' (a')^2 d_{ij}^{'-\alpha'} \left[ \sum\limits_{e=1}^{r^2} \left( \kappa_m \kappa_n \right)_e^{\beta/D} \right]^{\frac{D(\eta - 1) + \alpha'}{\beta}} \\
		&= \nu' (a')^2 d_{ij}^{'-\alpha'} \left[ \sum\limits_{e=1}^{r^2} \left( \frac{w_{mn}}{\nu a^2 d_{mn}^{-\alpha} } \right)_e^{\frac{\beta}{D(\eta - 1) + \alpha}} \right]^{\frac{D(\eta - 1) + \alpha'}{\beta}}. \\
	\end{aligned}
\end{equation}
In the last step, we have assumed that, for every pair of nodes $(m, n)$, we can obtain the product $\kappa_m \kappa_n$ from the corresponding weight $\omega_{mn}$, which is not true in general, as some links might not exist. However, this should be a reasonable approximation, since it only misses the smallest products of hidden degrees. Now, the above transformation cannot be performed without the precise distances in the embedding, as it depends on $d_{mn}$, but recalling that $d_{mn} = R \Delta \theta_{mn}$, where $\Delta \theta_{mn}$ stands for the angular separation between the nodes, and the fact that all such distances are approximately equal to the angular separation between the supernodes to which the nodes belong ($\Delta \theta_{mn} \approx \Delta \theta'_{ij}$), we can see that fixing $\alpha' = \alpha$ will remove all dependency on the distance,
\begin{equation}
	\begin{aligned}
		\omega'_{ij} &= \nu' (a')^2 d_{ij}^{'-\alpha} \left[ \sum\limits_{e=1}^{r^2} \left( \frac{w_{mn}}{\nu a^2 d_{mn}^{-\alpha} } \right)_e^{\frac{\beta}{D(\eta - 1) + \alpha}} \right]^{\frac{D(\eta - 1) + \alpha}{\beta}} \\
		&= \frac{\nu'}{\nu} \left( \frac{a'}{a} \right)^2 \left( \frac{R \Delta \theta'_{ij}}{R' \Delta \theta'_{ij}} \right)^\alpha \left[ \sum\limits_{e=1}^{r^2} \left( w_{mn} \right)_e^{\frac{\beta}{D(\eta - 1) + \alpha}} \right]^{\frac{D(\eta - 1) + \alpha}{\beta}} \\
		&= \frac{\nu'}{\nu} \left( \frac{a'}{a} \right)^2 r^{\alpha/D} \left[ \sum\limits_{e=1}^{r^2} \left( w_{mn} \right)_e^{\frac{\beta}{D(\eta - 1) + \alpha}} \right]^{\frac{D(\eta - 1) + \alpha}{\beta}},
	\end{aligned}
\end{equation}
where we have used that $R' = R/r^{1/D}$. 

Finally, we can choose any appropriate relation between primed and unprimed global parameters leading to
\begin{equation}
	\omega'_{ij} = C\left[ \sum\limits_{e=1}^{r^2} \left( w_{mn} \right)_e^{\phi} \right]^{1/\phi},
	\label{eq:transformation2}
\end{equation}
with $\phi \equiv \frac{\beta}{D(\eta - 1) + \alpha}$ and $C=\frac{\nu'}{\nu} \left( \frac{a'}{a} \right)^2 r^{\alpha/D}$. Therefore, the weighted model predicts that the exponent $\eta$ characterizing the relation between strength and degree is preserved in the renormalized network if weights are transformed following Eq.~(\ref{eq:transformation2}) (in the noiseless case) and the value of $\phi$ that corresponds to the considered network is used. 

\subsection{Theoretical derivation of the flow of the average strength}
\label{D}

We start from Eq.~(\ref{eq:transformation2}) ($D=1$) and impose that the rescaling variable $$C=\frac{\nu'}{\nu} \left( \frac{a'}{a} \right)^2 r^{\alpha}$$ is constant in the flow such that the transformation of weights keeps the same units in all scales of observation. The transformation of the relation between hidden strength and hidden degree is $$\frac{a'}{a}=\frac{\langle \sigma \rangle'}{\langle \sigma \rangle}\frac{\langle\kappa^\eta\rangle}{\langle\kappa^\eta\rangle'}.$$ We can also obtain the transformation of the free parameter $\nu$ using its expression from~\cite{allard2017geometric} and the expression for the parameter $\mu$, 
\begin{equation}
	\begin{aligned}
		\nu=\frac{\Gamma(1/2)}{2\pi^{1/2}\mu^{1-\alpha}I_2I_3 \langle\sigma\rangle}  & &\mu=\frac{\Gamma(1/2)}{2\pi^{1/2}I_1\langle k \rangle},
	\end{aligned}
	\nonumber
\end{equation}
which leads to
$$\frac{\nu'}{\nu}=\frac{\langle \sigma \rangle}{\langle \sigma \rangle'}\left(\frac{\langle k \rangle'}{\langle k \rangle}\right)^{1-\alpha}$$
and therefore to
$$C=\frac{\langle \sigma \rangle'}{\langle \sigma \rangle}\left(\frac{\langle\kappa^\eta\rangle}{\langle\kappa^\eta\rangle'} \right)^2 r^{\xi(1-\alpha)+\alpha},$$
where we have used the expression for the flow of the average degree. We use $\langle\kappa^\eta\rangle=\frac{\gamma-1}{\gamma-1-\eta}\kappa_0^{\eta}$ ($\eta<\gamma-1$) to compute its flow, and we obtain $$\frac{\langle\kappa^\eta\rangle'}{\langle\kappa^\eta\rangle}=r^{\xi \eta}.$$ Finally, $$C=\frac{\langle \sigma \rangle'}{\langle \sigma \rangle}r^{\xi(1-2\eta-\alpha)+\alpha}=\frac{\langle \sigma \rangle'}{\langle \sigma \rangle}r^{-\psi},$$ and we impose $C=1$ to obtain 
\begin{equation}
	\langle \sigma' \rangle=  \langle \sigma \rangle r^{\psi}, \mbox{      }\psi=\left(\frac{\alpha}{D}+2 \eta - 1\right)\xi - \frac{\alpha}{D},
	\label{eq:renors}
\end{equation}
from which $\psi>0$ implies an increasing average strength in the flow while it decreases if $\psi<0$.

\subsection{The transformation sum-GRW does not preserve the relation between strength and degree}
\label{E}
The sum-GRW transformation is
\begin{equation}\label{eq:wij2}
	\begin{aligned}
		w'_{ij} &=\sum\limits_{e=1}^{r^2} \epsilon_{mn} w_{mn} \\
		&=\nu d_{mn}^{-\alpha} \sum\limits_{e=1}^{r^2} \epsilon_{mn} \sigma_m \sigma_n ( \kappa_m \kappa_n )^{\alpha/D-1},
	\end{aligned}
\end{equation}
where $e$ runs over all pairs of nodes $(m, n)$ with $m$ in supernode $i$ and $n$ in supernode $j$ and $d_{mn} = R \Delta \theta_{mn}$, where $\Delta \theta_{mn}$ stands for the angular separation between the nodes. All such distances are approximately equal to the angular separation between the supernodes to which the nodes belong ($\Delta \theta_{mn} \approx \Delta \theta'_{ij}$), and one can take $\alpha' = \alpha$. Comparing Eq. (1) in main text and ~\eqref{eq:wij2}, we can write
\begin{equation}\label{eq:part1}
	\begin{aligned}
		\nu' d_{ij}^{'-\alpha'}=\nu d_{ij}^{-\alpha},
	\end{aligned}
\end{equation}
\begin{equation}\label{eq:part2}
	\begin{aligned}
		\epsilon_{ij}' \sigma_i' \sigma_j'  ( \kappa_i' \kappa_j' )^{\alpha'/D-1}&=\sum\limits_{e=1}^{r^2} \epsilon_{mn} \sigma_m \sigma_n ( \kappa_m \kappa_n )^{\alpha/D-1},
	\end{aligned}
\end{equation}
and using $R' = R/r^{1/D}$ and Eq.~\eqref{eq:part1} altogether, we have 
\begin{equation}\label{eq:nu}
	\begin{aligned}
		\nu'=\nu r^{-\alpha/D}.
	\end{aligned}
\end{equation}
Therefore, in the noiseless version ($\epsilon_{mn} = 1 \; \forall (m,n)$), 
we can obtain the hidden strength $\sigma_i'$ in the supernodes layer as
\begin{equation}\label{eq:sigma}
	\begin{aligned}
		\sigma_i' =\frac{\sum\limits_{m=1}^{r}\sigma_m \kappa_m ^{\alpha/D-1}}{\kappa_i'^{\alpha/D-1}}\not\approx  \kappa_i'^\eta,
	\end{aligned}
\end{equation}
which proves that, in general, the relation between hidden strength and hidden degree is not preserved under sum-GRW.

\subsection{Scale down replicas}
\label{replicas}
\begin{enumerate}
	\item
	We obtain a renormalized network layer by applying the sup-GRW method with a given value of $r$ and number of iterations to match the target network size.
	\item
	Typically, the average degree of the renormalized network layer is higher than the original one. Thus, to obtain a scaled down network replica of the topology, we decrease the average degree in the renormalized layer to that in the original network as explained in Ref.~\cite{Garcia2018}, such that $\langle k_{\mathrm{new}}^{(l)}\rangle=\langle k^{(0)}\rangle$. The main idea is to reduce the value of $\mu^{(l)}$ to a new
	one $\mu_{new}^{(l)}$, which means that the connection probability of every pair of nodes $(i,j)$, $p_{ij}^{(l)}$ decreases to $p_{ij, new}^{(l)}$. Therefore, the probability for a link to exist in the pruned network reads:
	\begin{equation}\label{eq:pij_new}
		\begin{aligned}
			p_{ij,new}^{(l)}=\frac{1}{1+\left( \frac{d_{ij}}{\mu_{new}^{(l)}\kappa_i\kappa_j}\right) ^\beta}.
		\end{aligned}
	\end{equation}
	In particular, we prune the links using $\mu_{\mathrm{new}}^{(l)}=h \frac{\langle k^{(0)}\rangle}{\langle k^{(l)}\rangle}\mu^{(l)}$ with $h=1$ as initial value. After an iteration for all the links in the layer, we give $h$ a new value $h(1-0.1u) \to h$ if $\langle k_{\mathrm{new}}^{(l)}\rangle> \langle k^{(0)}\rangle$, where $u \in [0,1)$ is a random variable from a uniform distribution. If $\langle k_{\mathrm{new}}^{(l)}\rangle< \langle k^{(0)}\rangle$, $h(1+0.1u) \to h$. The procedure stops when $|\langle k_{\mathrm{new}}^{(l)}\rangle- \langle k^{(0)}\rangle|$ is below a given threshold, that we set to $0.1$. 
	\item
	Finally, we rescale the weights in the resulting network by a global factor to match the average weight of the original network. Specifically, we calculate the average weight $\langle w_{\mathrm{new}}^{(l)}\rangle$ of the resulting network from step (2) and the average weight $\langle w^{(0)}\rangle$ in the original network. Then we rescale the weight of each link by the factor $c=\frac{\langle w^{(0)}\rangle}{\langle w_{\mathrm{new}}^{(l)}\rangle}$. 
\end{enumerate}

\clearpage
\newpage
\section{Results for sup-GRW}
\subsection{sup-GRW in empirical data}
\begin{figure}[ht]
	\includegraphics[width=0.8\linewidth]{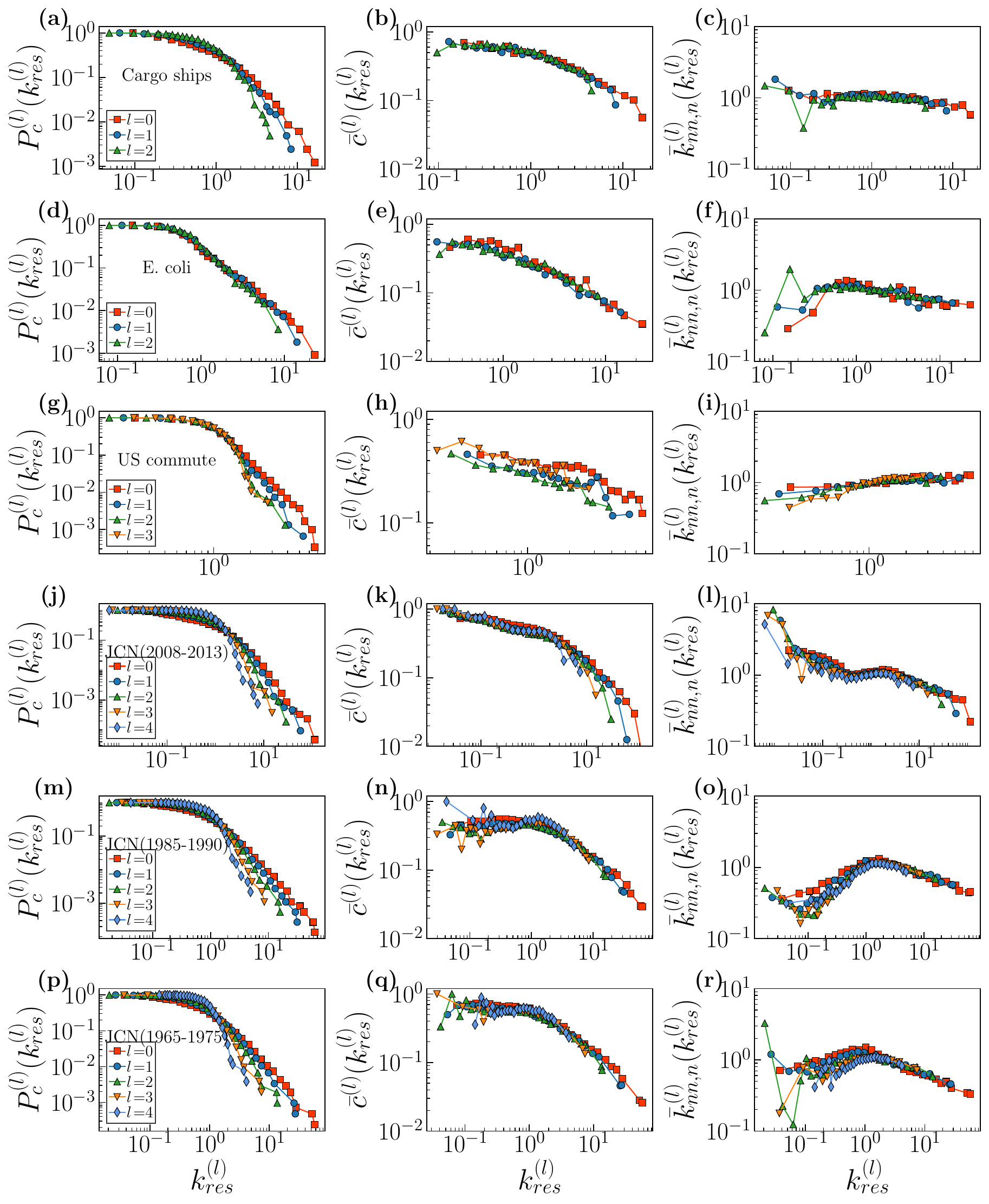}
	\caption{\textbf{Network properties for sup-GRW in different empirical networks.} First column: complementary cumulative degree distribution $P_c^{(l)}(k_{res}^{(l)})$, Second column: the degree-dependent clustering coefficient $\bar{c}^{(l)}(k_{res}^{(l)})$, Last column:  normalized average nearest-neighbour degree $\bar{k}_{nn,n}^{(l)} (k_{res}^{(l)}) = \bar{k}_{nn}^{(l)} (k_{res}^{(l)}) \langle  k^{(l)}\rangle/\langle(k^{(l)})^2\rangle$ of rescaled degrees $k^{(l)}_{res}$ for different layers $l$. Each row indicates an empirical network.}
\end{figure}

\begin{figure}[ht] 	
	\includegraphics[width=0.8\linewidth]{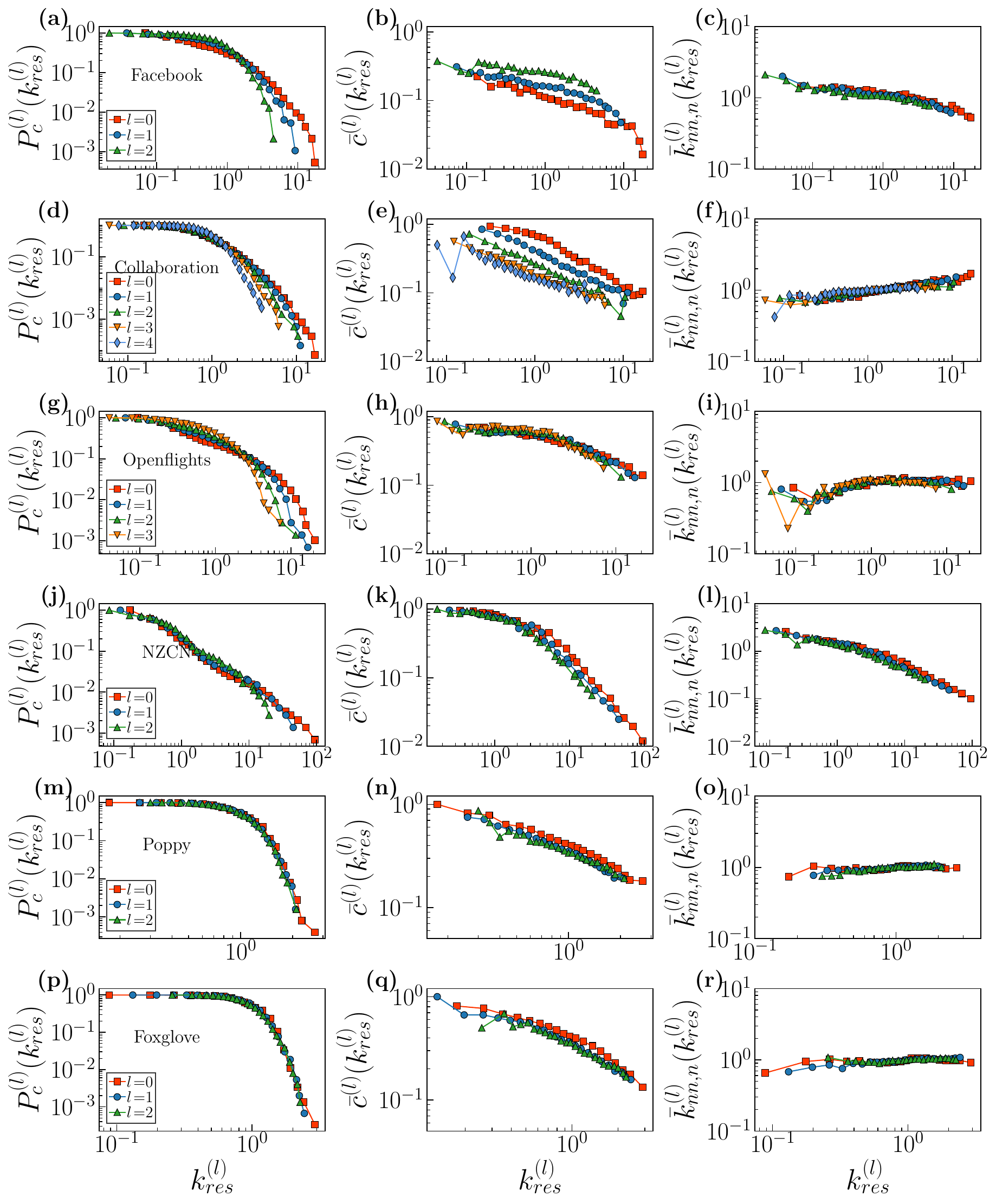} 
	\caption{\textbf{Network properties for sup-GRW in different empirical networks.} First column: complementary cumulative degree distribution $P_c^{(l)}(k_{res}^{(l)})$, Second column: the degree-dependent clustering coefficient $\bar{c}^{(l)}(k_{res}^{(l)})$, Last column:  normalized average nearest-neighbour degree $\bar{k}_{nn,n}^{(l)} (k_{res}^{(l)}) = \bar{k}_{nn}^{(l)} (k_{res}^{(l)}) \langle  k^{(l)}\rangle/\langle(k^{(l)})^2\rangle$ of rescaled degrees $k^{(l)}_{res}$ for different layers $l$. Each row indicates an empirical network.}
\end{figure}

\begin{figure}[ht] 	
	\includegraphics[width=1\linewidth]{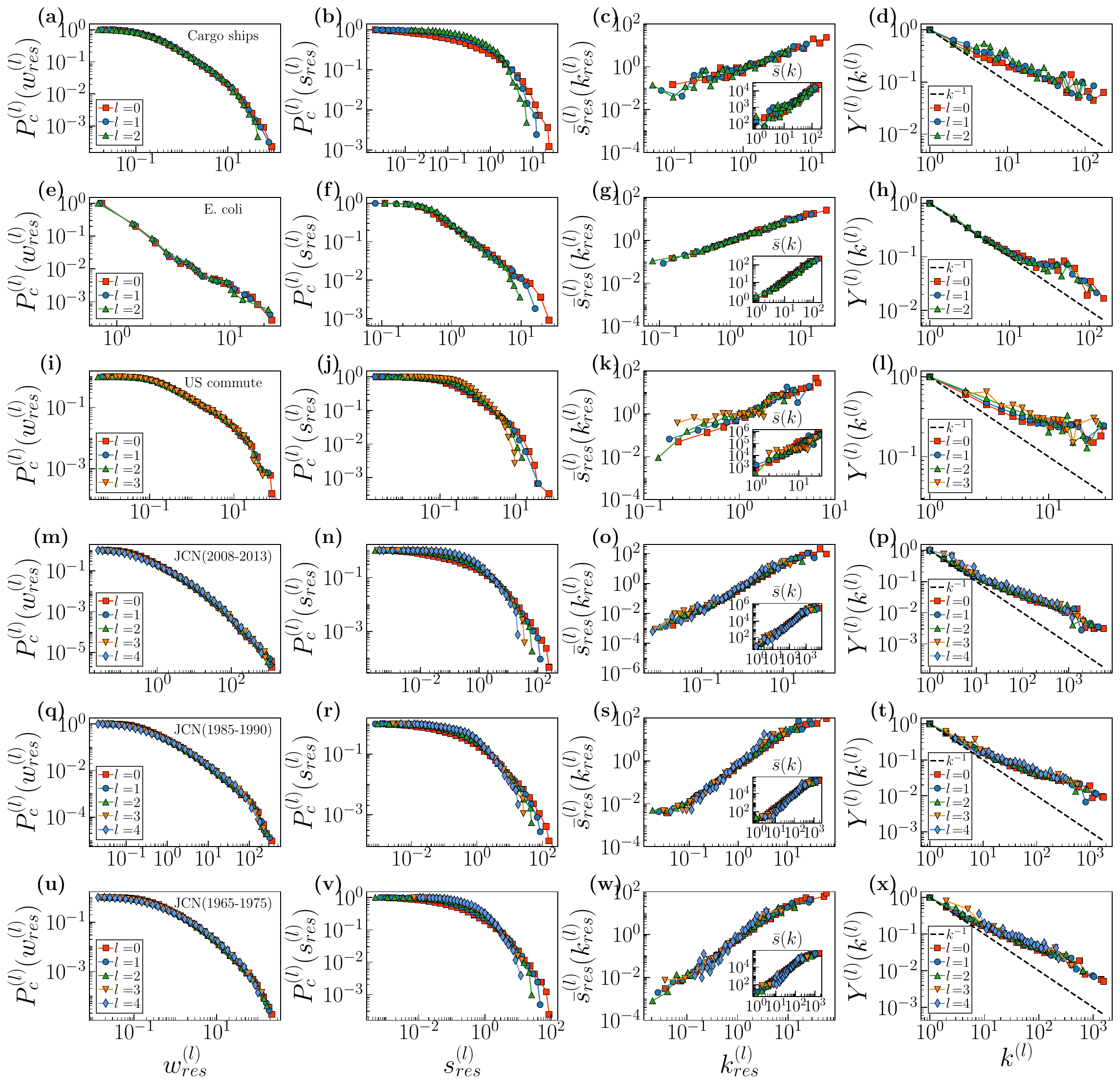} 
	\caption{\textbf{Network properties for sup-GRW in different empirical networks.} First column: complementary cumulative weight distributions $P_c^{(l)}(w_{res}^{(l)})$ of rescaled weights $w^{(l)}_{res}=w^{(l)}/\langle w^{(l)}\rangle$ for different layers $l$.
		Second column: complementary cumulative strength distributions $P_c^{(l)}(s_{res}^{(l)})$ of rescaled strengths $s^{(l)}_{res}=s^{(l)}/\langle s^{(l)}\rangle$ for different layers $l$.
		Third column: average rescaled strengths as a function of rescaled degrees, i.e.,  $\bar{s}_{res}^{(l)}(k_{res}^{(l)})=\bar{s}^{(l)}(k)/\langle s\rangle^{(l)}$. Inset shows average strength as a function of degree.
		Last column: disparity of nodes as a function of their degree. Each row indicates an empirical network.}
\end{figure}

\begin{figure}[ht] 	
	\includegraphics[width=1\linewidth]{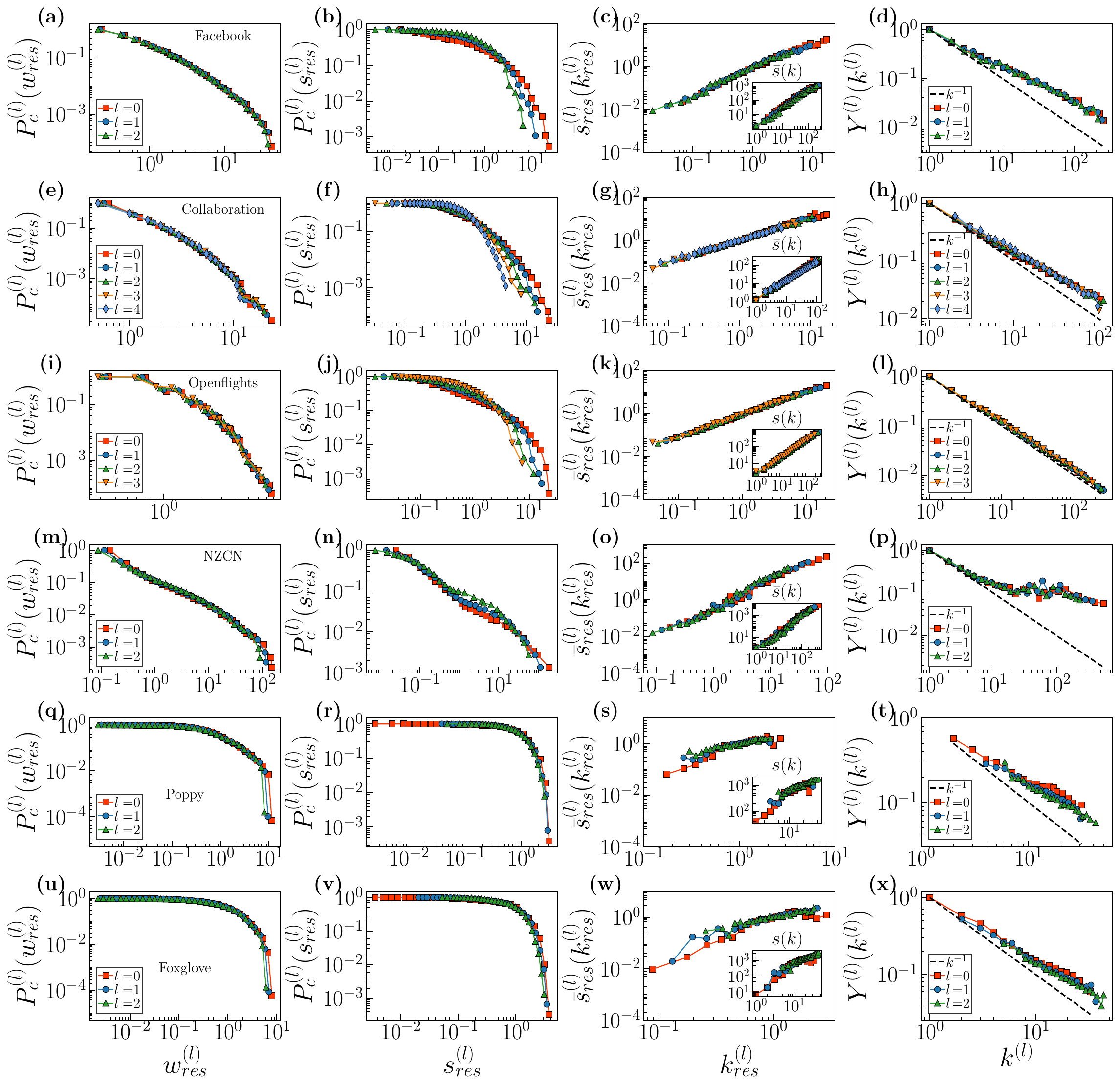} 
	\caption{\textbf{Network properties for sup-GRW in different empirical networks.} First column: complementary cumulative weight distributions $P_c^{(l)}(w_{res}^{(l)})$ of rescaled weights $w^{(l)}_{res}=w^{(l)}/\langle w^{(l)}\rangle$ for different layers $l$.
		Second column: complementary cumulative strength distributions $P_c^{(l)}(s_{res}^{(l)})$ of rescaled strengths $s^{(l)}_{res}=s^{(l)}/\langle s^{(l)}\rangle$ for different layers $l$.
		Third column: average rescaled strengths as a function of rescaled degrees, i.e.,  $\bar{s}_{res}^{(l)}(k_{res}^{(l)})=\bar{s}^{(l)}(k)/\langle s\rangle^{(l)}$. Inset shows average strength as a function of degree.
		Last column: disparity of nodes as a function of their degree. Each row indicates an empirical network.}
\end{figure}

\begin{figure}[ht] 
	\includegraphics[width=0.58\linewidth]{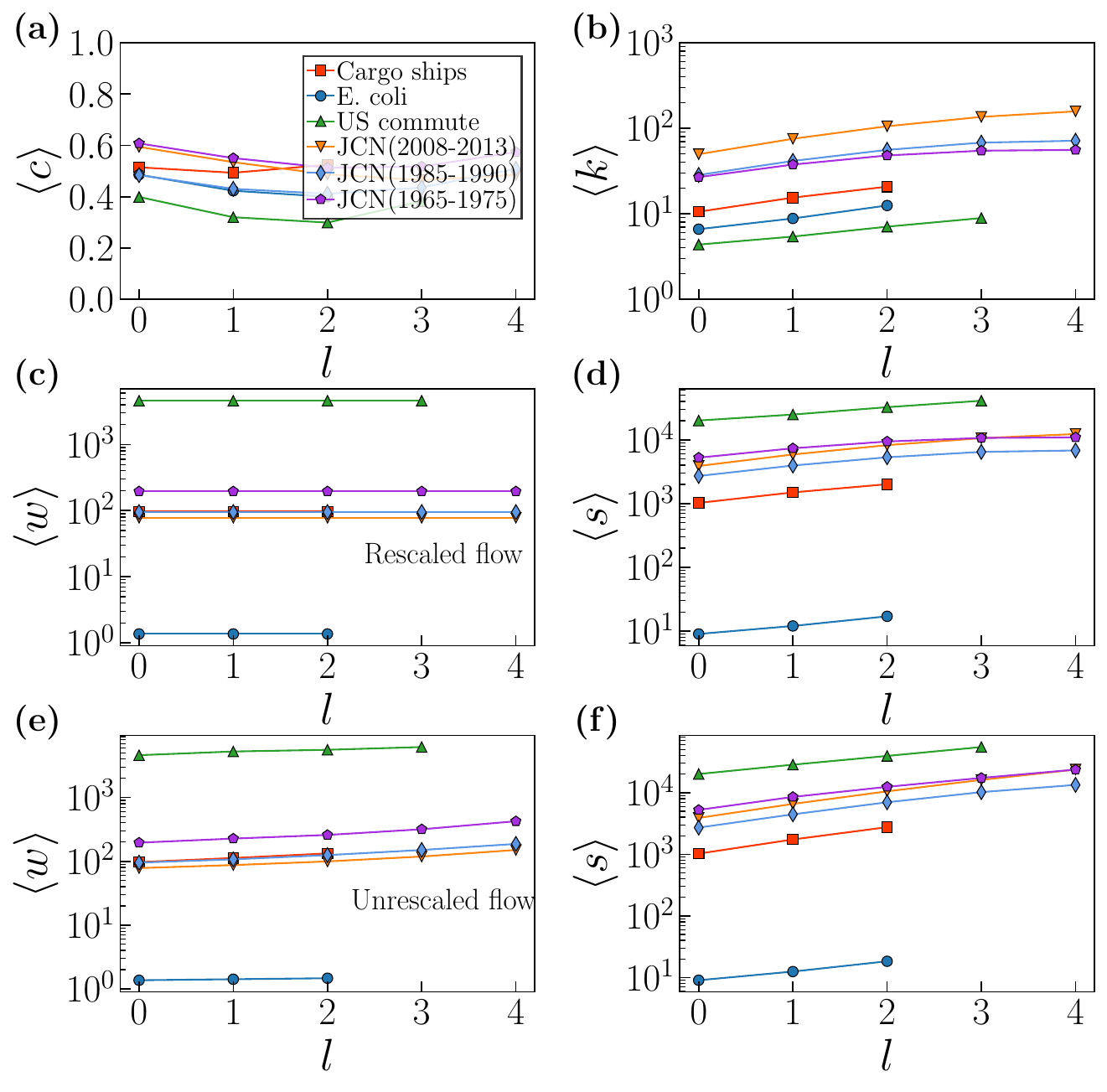}
	\caption{(a) average clustering coefficient, (b) average degree, (c) rescaled average weight and (d) corresponding average strength, (e) unrescaled average weight and (f) corresponding average strength for different layers $l$. }
\end{figure}

\begin{figure}[ht] 
	\includegraphics[width=0.58\linewidth]{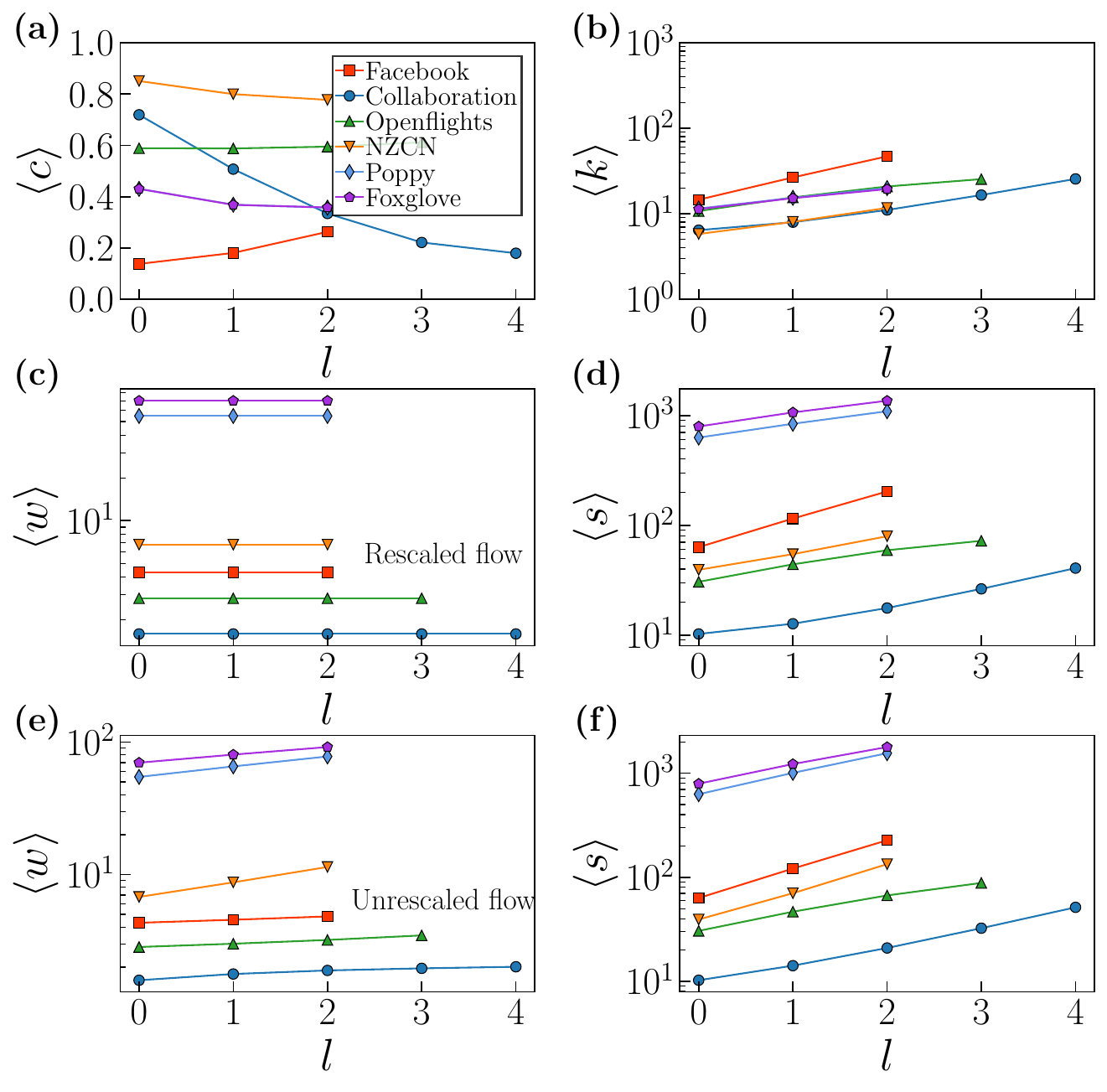}
	\caption{(a) average clustering coefficient, (b) average degree, (c) rescaled average weight and (d) corresponding average strength, (e) unrescaled average weight and (f) corresponding average strength for different layers $l$. }
\end{figure}

\clearpage
\newpage

\subsection{Semigroup structure in sup-GRW transformation} 
\begin{figure}[ht] 	
	\includegraphics[width=0.9\linewidth]{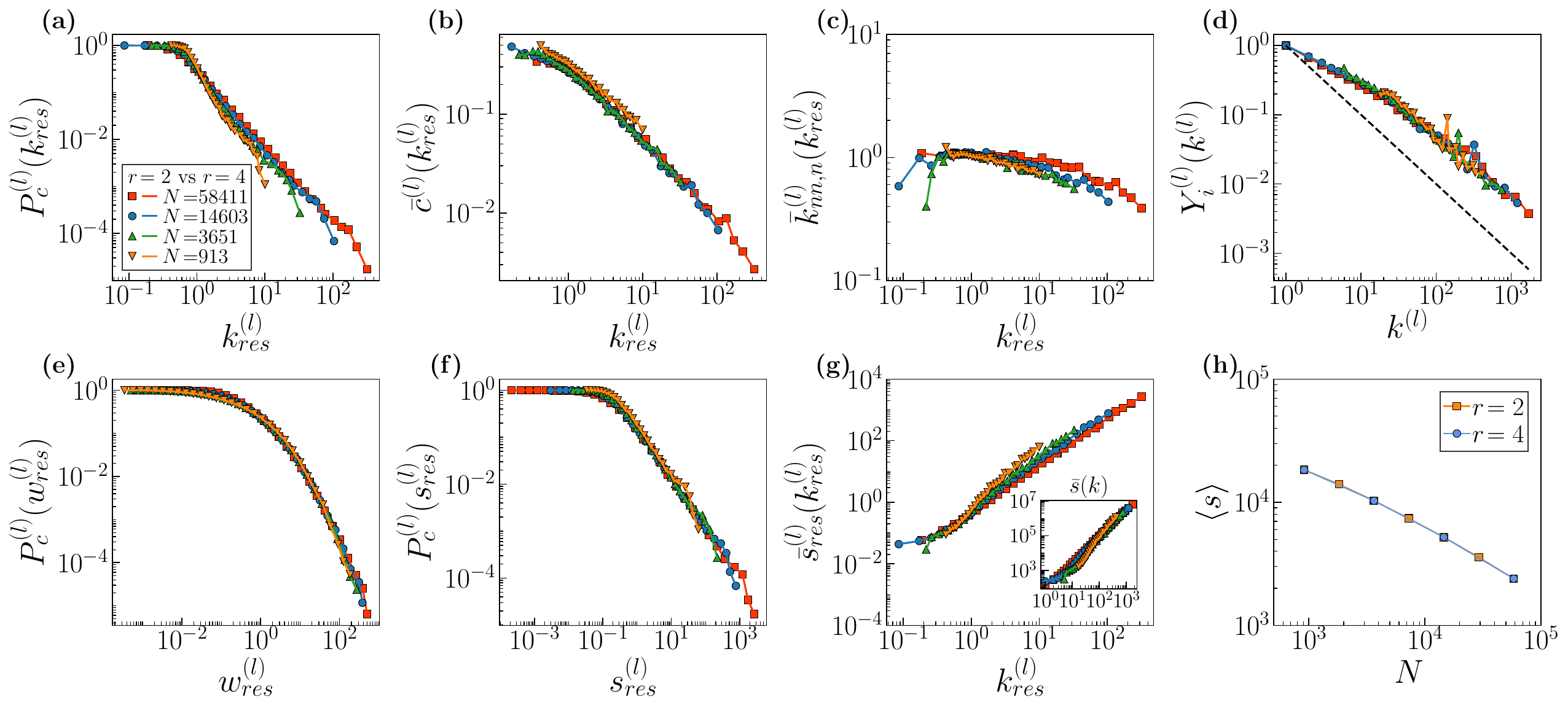}
	\caption{\textbf{Semigroup structure of sup-GRW in a synthetic network with $r=2$ and $r=4$.} (a) Complementary cumulative degree distribution $P_c^{(l)}(k_{res}^{(l)})$, (b) The degree-dependent clustering coefficient $\bar{c}^{(l)}(k_{res}^{(l)})$, (c) Normalized average nearest-neighbour degree $\bar{k}_{nn,n}^{(l)} (k_{res}^{(l)}) = \bar{k}_{nn}^{(l)} (k_{res}^{(l)}) \langle  k^{(l)}\rangle/\langle(k^{(l)})^2\rangle$ of rescaled degrees $k^{(l)}_{res}$ for different layers $l$. (d) Disparity of nodes as a function of their degree. 
		(e) Complementary cumulative weight distributions $P_c^{(l)}(w_{res}^{(l)})$ of rescaled weights $w^{(l)}_{res}=w^{(l)}/\langle w^{(l)}\rangle$ for different layers $l$. (f) Complementary cumulative strength distributions $P_c^{(l)}(s_{res}^{(l)})$ of rescaled strengths $s^{(l)}_{res}=s^{(l)}/\langle s^{(l)}\rangle$ for different layers $l$. (g) average rescaled strengths as a function of rescaled degrees, i.e.,  $\bar{s}_{res}^{(l)}(k_{res}^{(l)})=\bar{s}^{(l)}(k)/\langle s\rangle^{(l)}$. Inset shows average strength as a function of degree. (h) Average strength $\langle s\rangle$ as a function of the network size $N$. The parameters are $N=58411$, $\beta= 1.5$, $\mu=0.0413$, $a=100$, $\eta=1.5$, $\alpha=0.40$, $\gamma=2.5$, and $\langle \epsilon^2 \rangle=1.0$, $\phi=1.67$. Here symbols show the case of $r=2$ and lines are the one of $r=4$.}  
	\label{fig:semigroup}
\end{figure}

The geometric renormalization transformation has Abelian semigroup structure with respect to the composition, meaning that a certain number of iterations of a given resolution are equivalent to a single transformation of higher resolution.  
We here validated the semigroup structure in sup-GRW transformation with  synthetic and empirical networks. Given an original network, we performed the sup-GRW
with $r=2$ and $r=4$, respectively. When the geometric renormalization transformation to the same network size, we compared the their network properties. Figure~\ref{fig:semigroup} shows the results for a representative synthetic network. 

\clearpage
\newpage
\section{Results for sum-GRW}

\begin{figure}[ht] 	
	\includegraphics[width=1\linewidth]{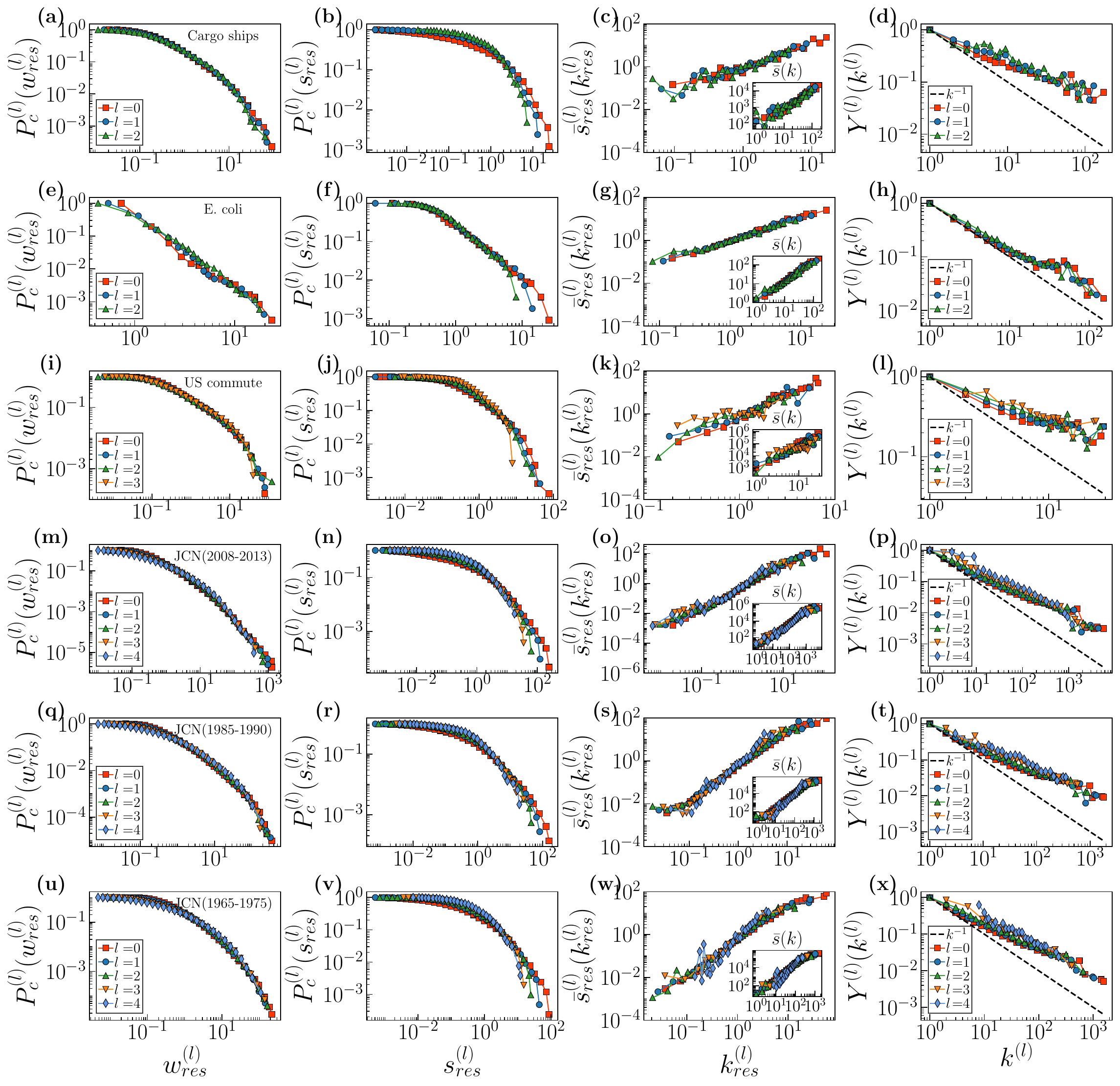} 
	\caption{\textbf{Network properties for sum-GRW in different empirical networks.} First column: complementary cumulative weight distributions $P_c^{(l)}(w_{res}^{(l)})$ of rescaled weights $w^{(l)}_{res}=w^{(l)}/\langle w^{(l)}\rangle$ for different layers $l$.
		Second column: complementary cumulative strength distributions $P_c^{(l)}(s_{res}^{(l)})$ of rescaled strengths $s^{(l)}_{res}=s^{(l)}/\langle s^{(l)}\rangle$ for different layers $l$.
		Third column: average rescaled strengths as a function of rescaled degrees, i.e.,  $\bar{s}_{res}^{(l)}(k_{res}^{(l)})=\bar{s}^{(l)}(k)/\langle s\rangle^{(l)}$. Inset shows average strength as a function of degree.
		Last column: disparity of nodes as a function of their degree. Each row indicates an empirical network.}
\end{figure}

\begin{figure}[ht] 	
	\includegraphics[width=1\linewidth]{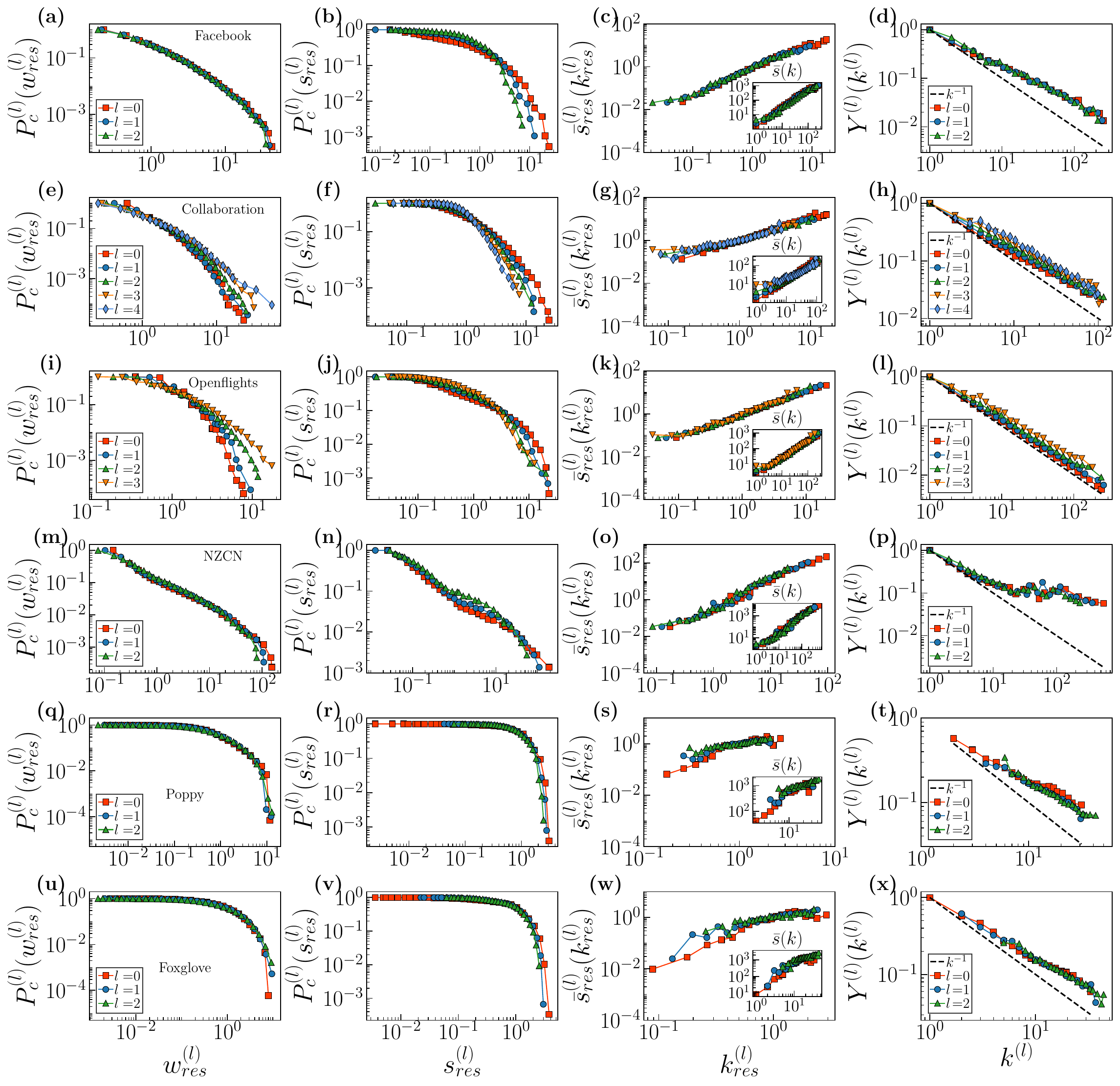} 
	\caption{\textbf{Network properties for sum-GRW in different empirical networks.} First column: complementary cumulative weight distributions $P_c^{(l)}(w_{res}^{(l)})$ of rescaled weights $w^{(l)}_{res}=w^{(l)}/\langle w^{(l)}\rangle$ for different layers $l$.
		Second column: complementary cumulative strength distributions $P_c^{(l)}(s_{res}^{(l)})$ of rescaled strengths $s^{(l)}_{res}=s^{(l)}/\langle s^{(l)}\rangle$ for different layers $l$.
		Third column: average rescaled strengths as a function of rescaled degrees, i.e.,  $\bar{s}_{res}^{(l)}(k_{res}^{(l)})=\bar{s}^{(l)}(k)/\langle s\rangle^{(l)}$. Inset shows average strength as a function of degree.
		Last column: disparity of nodes as a function of their degree. Each row indicates an empirical network.}
\end{figure} 

\clearpage
\newpage
\section{Results for $\phi$-GRW}
\subsection{$\phi$-GRW in empirical data}

%

\begin{figure}[ht] 	
	\includegraphics[width=1\linewidth]{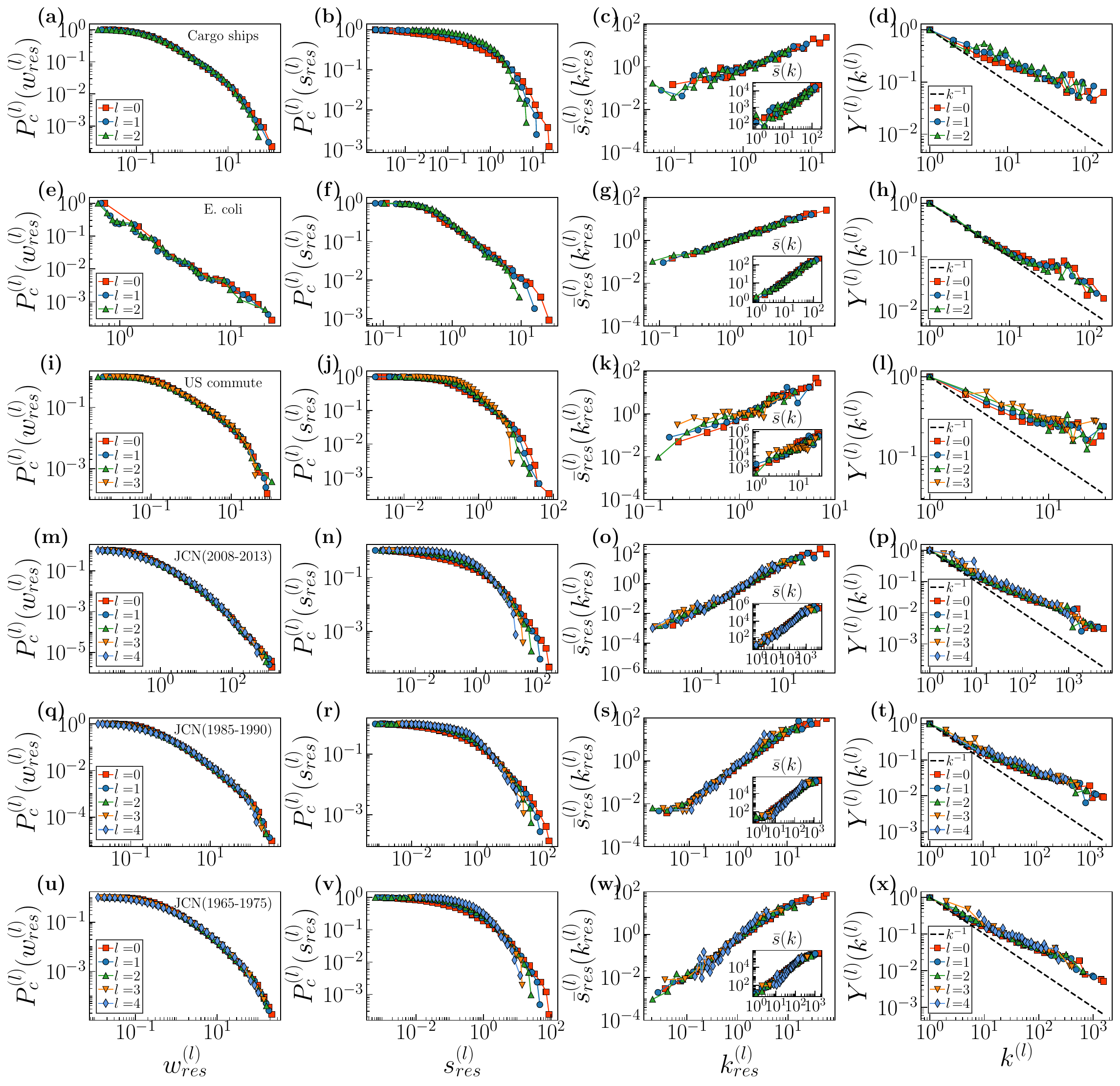} 
	\caption{\textbf{Network properties for $\phi$-GRW in different empirical networks.} First column: complementary cumulative weight distributions $P_c^{(l)}(w_{res}^{(l)})$ of rescaled weights $w^{(l)}_{res}=w^{(l)}/\langle w^{(l)}\rangle$ for different layers $l$.
		Second column: complementary cumulative strength distributions $P_c^{(l)}(s_{res}^{(l)})$ of rescaled strengths $s^{(l)}_{res}=s^{(l)}/\langle s^{(l)}\rangle$ for different layers $l$.
		Third column: average rescaled strengths as a function of rescaled degrees, i.e.,  $\bar{s}_{res}^{(l)}(k_{res}^{(l)})=\bar{s}^{(l)}(k)/\langle s\rangle^{(l)}$. Inset shows average strength as a function of degree.
		Last column: disparity of nodes as a function of their degree. Each row indicates an empirical network.}
\end{figure}

\begin{figure}[ht] 	
	\includegraphics[width=1\linewidth]{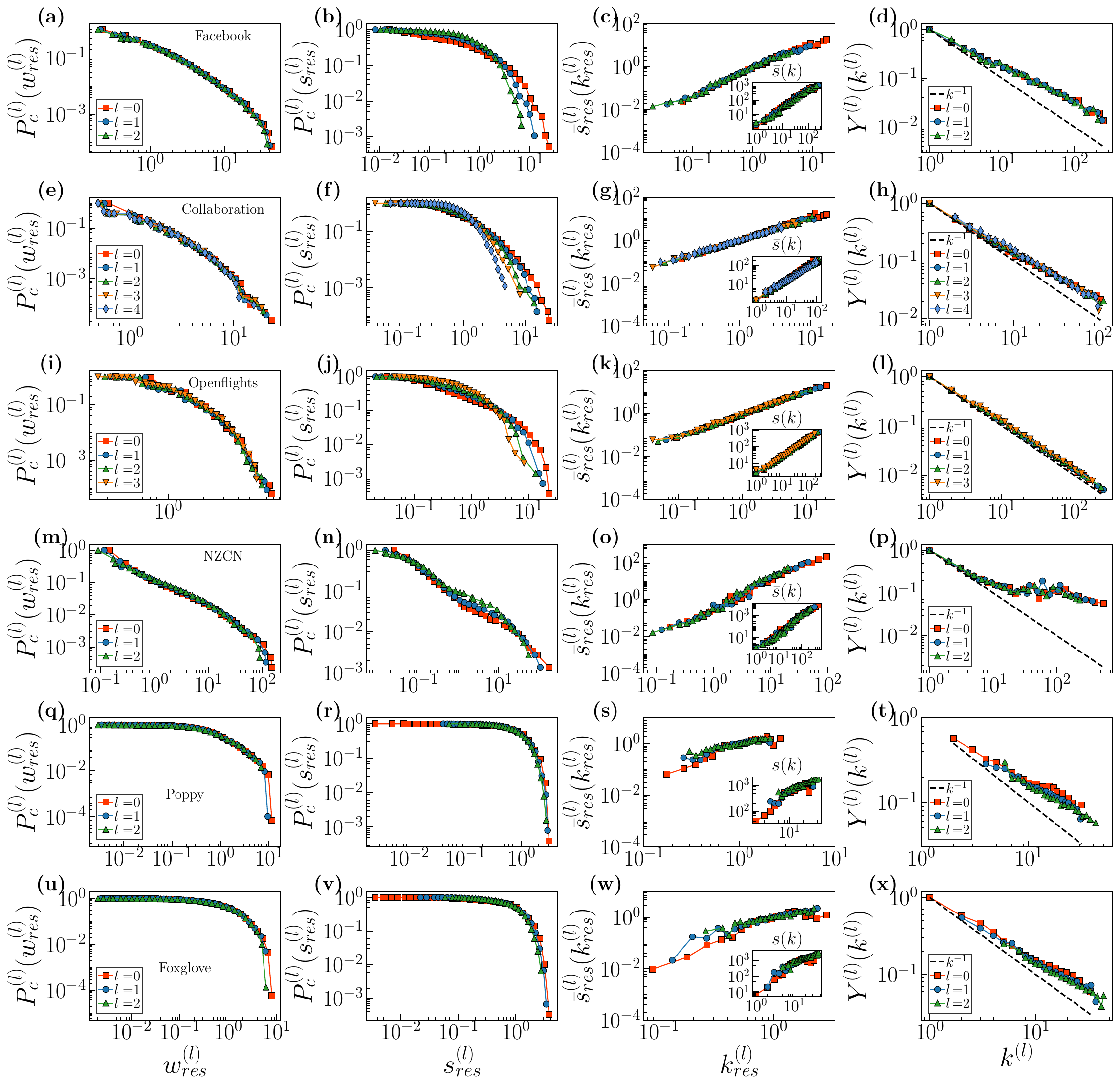} 
	\caption{\textbf{Network properties for $\phi$-GRW in different empirical networks.} First column: complementary cumulative weight distributions $P_c^{(l)}(w_{res}^{(l)})$ of rescaled weights $w^{(l)}_{res}=w^{(l)}/\langle w^{(l)}\rangle$ for different layers $l$.
		Second column: complementary cumulative strength distributions $P_c^{(l)}(s_{res}^{(l)})$ of rescaled strengths $s^{(l)}_{res}=s^{(l)}/\langle s^{(l)}\rangle$ for different layers $l$.
		Third column: average rescaled strengths as a function of rescaled degrees, i.e.,  $\bar{s}_{res}^{(l)}(k_{res}^{(l)})=\bar{s}^{(l)}(k)/\langle s\rangle^{(l)}$. Inset shows average strength as a function of degree.
		Last column: disparity of nodes as a function of their degree. Each row indicates an empirical network.}
\end{figure} 

\clearpage
\newpage
\subsection{$\phi-$GRW in synthetic networks}
\subsubsection{Semigroup structure}
\begin{figure}[ht] 
	\includegraphics[width=1\linewidth]{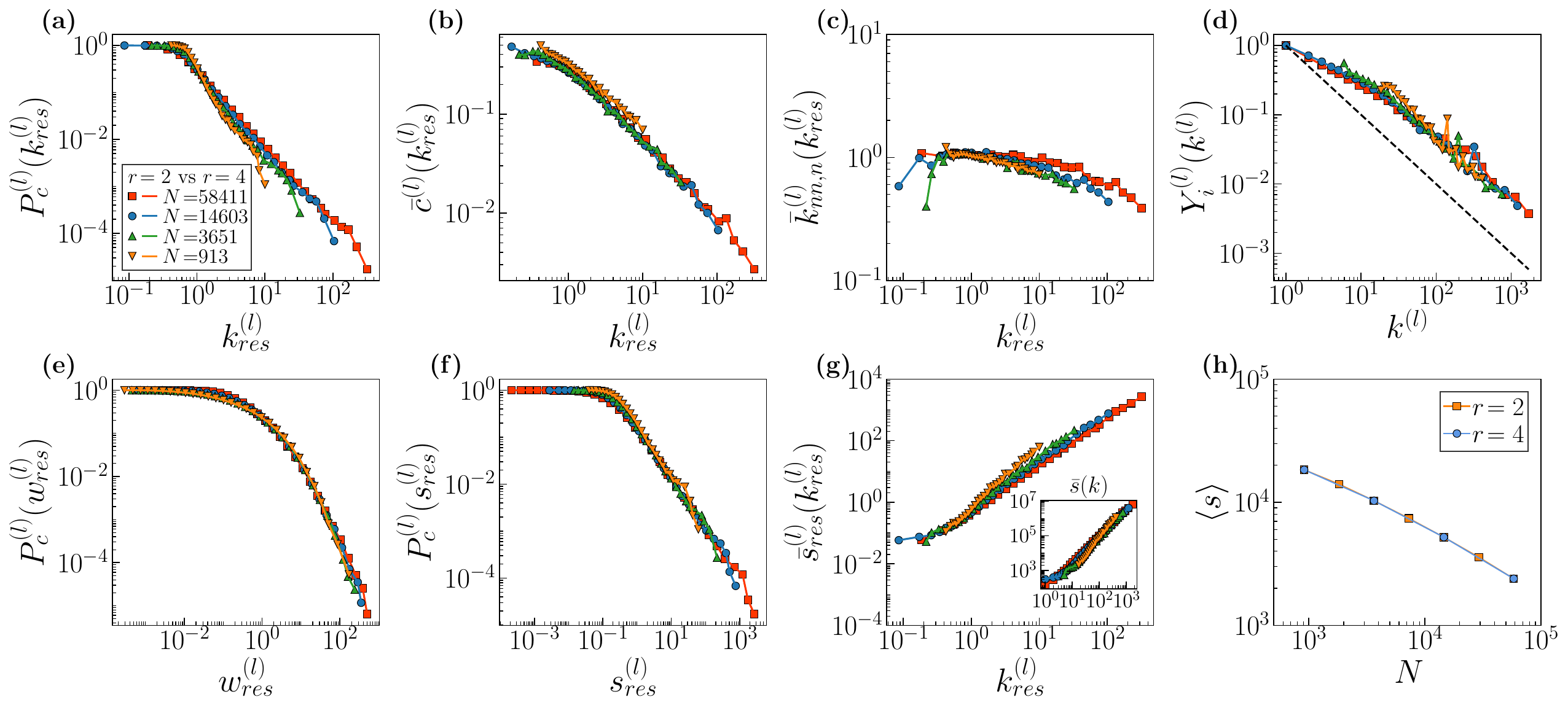}
	\caption{\textbf{Semigroup structure of $\phi-$GRW in synthetic network with $r=2$ and $r=4$}(a) Complementary cumulative degree distribution $P_c^{(l)}(k_{res}^{(l)})$, (b) The degree-dependent clustering coefficient $\bar{c}^{(l)}(k_{res}^{(l)})$, (c) Normalized average nearest-neighbour degree $\bar{k}_{nn,n}^{(l)} (k_{res}^{(l)}) = \bar{k}_{nn}^{(l)} (k_{res}^{(l)}) \langle  k^{(l)}\rangle/\langle(k^{(l)})^2\rangle$ of rescaled degrees $k^{(l)}_{res}$ for different layers $l$. (d) Disparity of nodes as a function of their degree. 
		(e) Complementary cumulative weight distributions $P_c^{(l)}(w_{res}^{(l)})$ of rescaled weights $w^{(l)}_{res}=w^{(l)}/\langle w^{(l)}\rangle$ for different layers $l$. (f) Complementary cumulative strength distributions $P_c^{(l)}(s_{res}^{(l)})$ of rescaled strengths $s^{(l)}_{res}=s^{(l)}/\langle s^{(l)}\rangle$ for different layers $l$. (g) average rescaled strengths as a function of rescaled degrees, i.e.,  $\bar{s}_{res}^{(l)}(k_{res}^{(l)})=\bar{s}^{(l)}(k)/\langle s\rangle^{(l)}$. Inset shows average strength as a function of degree. (h) Average strength $\langle s\rangle$ as a function of the network size $N$. The parameters are $N=58411$, $\beta= 1.5$, $\mu=0.0413$, $a=100$, $\eta=1.5$, $\alpha=0.40$, $\gamma=2.5$, and $\langle \epsilon^2 \rangle=1.0$, $\phi=1.67$. Here symbols show the case of $r=2$ and lines are the one of $r=4$.		
	}  
\end{figure}

\clearpage
\newpage
\subsubsection{The influence of $\alpha$}
\begin{figure}[ht] 
	\includegraphics[width=1\linewidth]{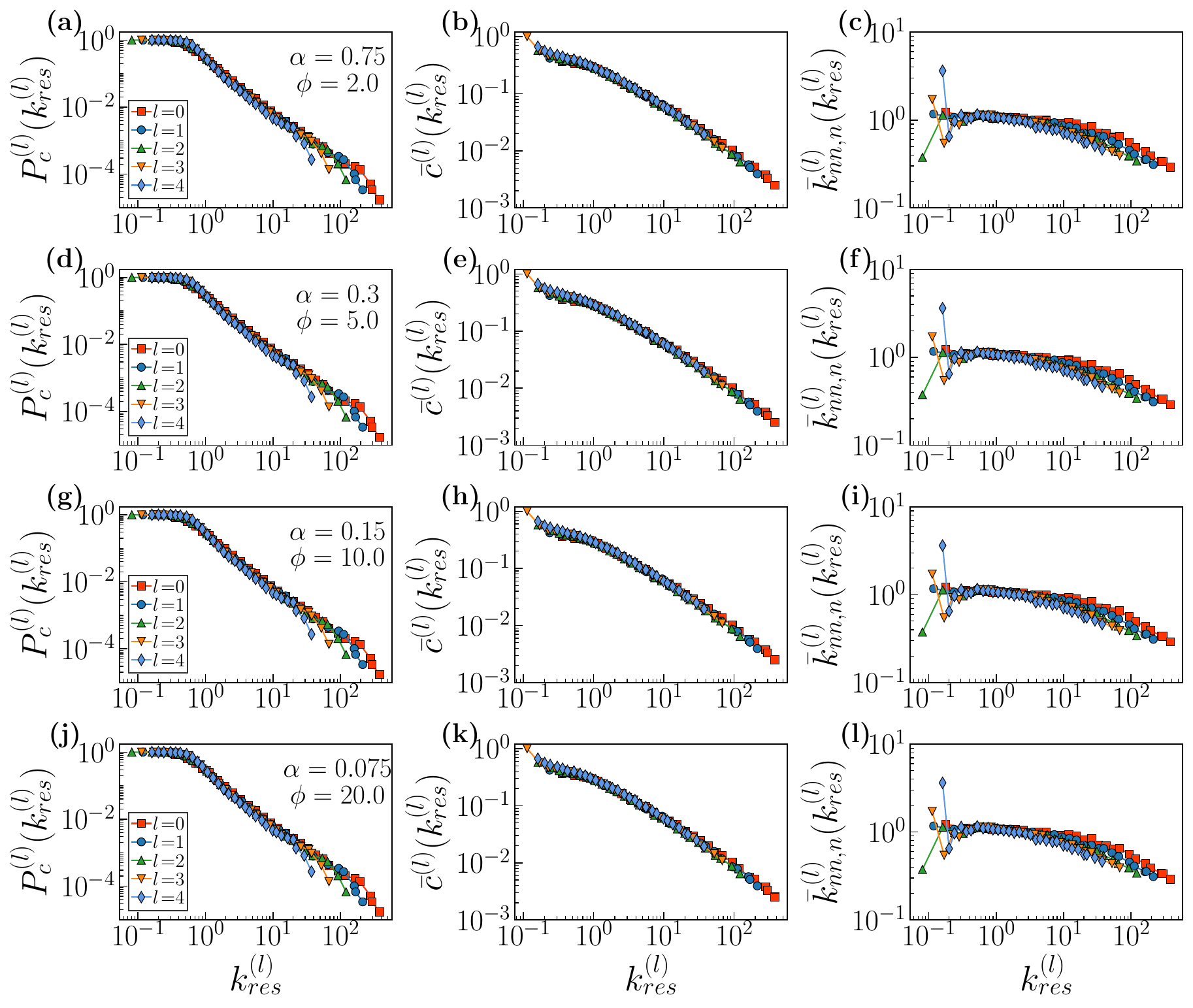}
	\caption{\textbf{$\phi$-GRW in synthetic network with different $\alpha$.} First column: complementary cumulative degree distribution $P_c^{(l)}(k_{res}^{(l)})$, Second column: the degree-dependent clustering coefficient $\bar{c}^{(l)}(k_{res}^{(l)})$, Last column:  normalized average nearest-neighbour degree $\bar{k}_{nn,n}^{(l)} (k_{res}^{(l)}) = \bar{k}_{nn}^{(l)} (k_{res}^{(l)}) \langle  k^{(l)}\rangle/\langle(k^{(l)})^2\rangle$ of rescaled degrees $k^{(l)}_{res}$ for different layers $l$. Each row indicates a synthetic  network. The parameters are $N=58948$, $\beta= 1.5$, $\mu=0.0413$, $a=100$, $\eta=1.0$, $\gamma=2.5$ and $\langle \epsilon^2 \rangle=1.0$.}  
\end{figure}
\begin{figure}[ht] 
	\includegraphics[width=1\linewidth]{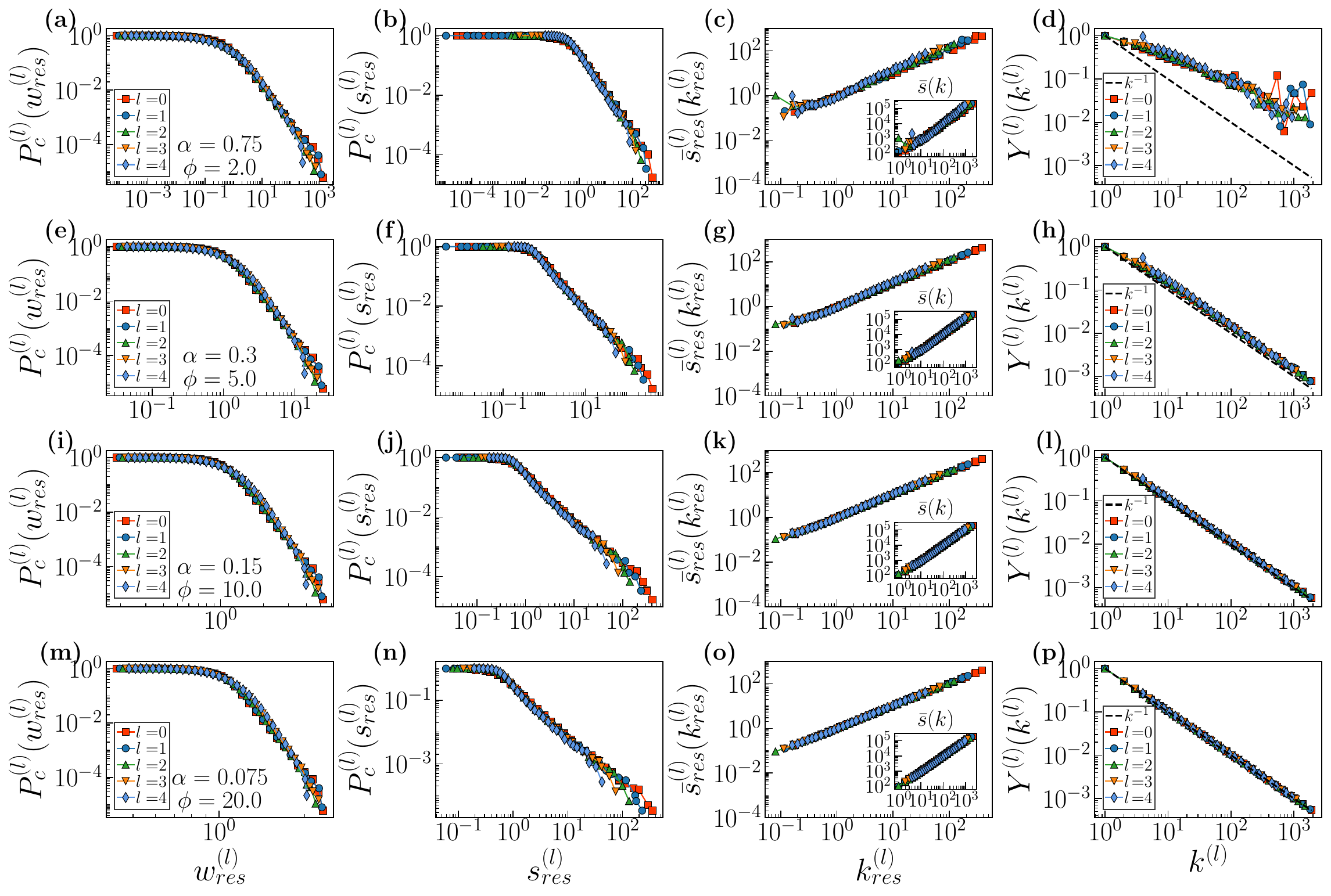}
	\caption{\textbf{$\phi$-GRW in synthetic network with different  $\alpha$.} First column: complementary cumulative weight distributions $P_c^{(l)}(w_{res}^{(l)})$ of rescaled weights $w^{(l)}_{res}=w^{(l)}/\langle w^{(l)}\rangle$ for different layers $l$.
		Second column: complementary cumulative strength distributions $P_c^{(l)}(s_{res}^{(l)})$ of rescaled strengths $s^{(l)}_{res}=s^{(l)}/\langle s^{(l)}\rangle$ for different layers $l$.
		Third column: average rescaled strengths as a function of rescaled degrees, i.e.,  $\bar{s}_{res}^{(l)}(k_{res}^{(l)})=\bar{s}^{(l)}(k)/\langle s\rangle^{(l)}$. Inset shows average strength as a function of degree.
		Last column: disparity of nodes as a function of their degree. Each row indicates a synthetic network.The parameters are $N=58948$, $\beta= 1.5$, $\mu=0.0413$, $a=100$, $\eta=1.0$, $\gamma=2.5$ and $\langle \epsilon^2 \rangle=1.0$.}  
\end{figure}

\clearpage
\newpage
\subsubsection{The influence of $\eta$}
\begin{figure}[ht] 
	\includegraphics[width=1\linewidth]{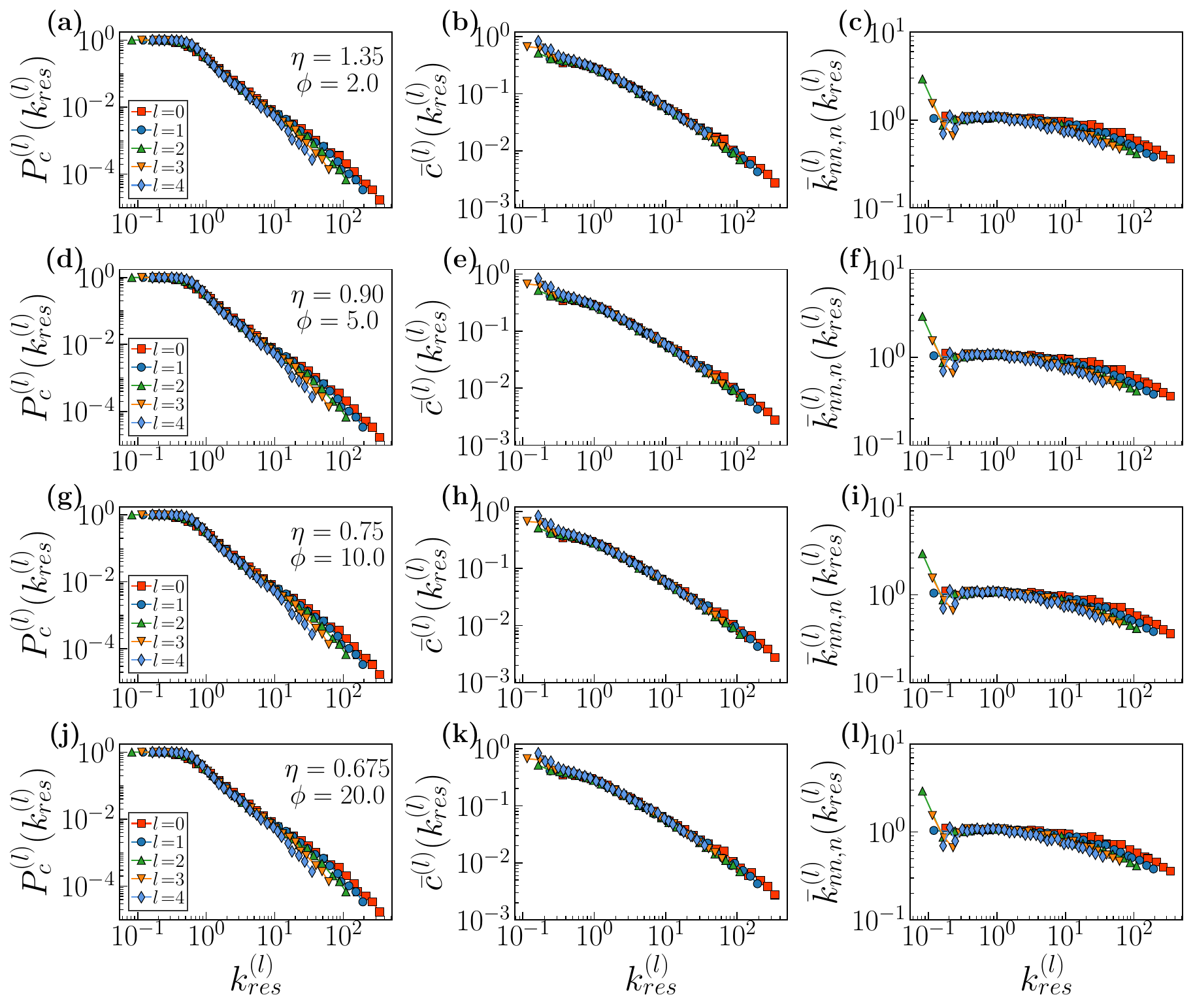}
	\caption{\textbf{$\phi$-GRW in synthetic network with different $\eta$.} First column: complementary cumulative degree distribution $P_c^{(l)}(k_{res}^{(l)})$, Second column: the degree-dependent clustering coefficient $\bar{c}^{(l)}(k_{res}^{(l)})$, Last column:  normalized average nearest-neighbour degree $\bar{k}_{nn,n}^{(l)} (k_{res}^{(l)}) = \bar{k}_{nn}^{(l)} (k_{res}^{(l)}) \langle  k^{(l)}\rangle/\langle(k^{(l)})^2\rangle$ of rescaled degrees $k^{(l)}_{res}$ for different layers $l$. Each row indicates a synthetic  network. The parameters are $N=58948$, $\beta= 1.5$, $\mu=0.0413$, $a=100$, $\eta=1.0$, $\gamma=2.5$ and $\langle \epsilon^2 \rangle=1.0$.}  
\end{figure}
\begin{figure}[ht] 
	\includegraphics[width=1\linewidth]{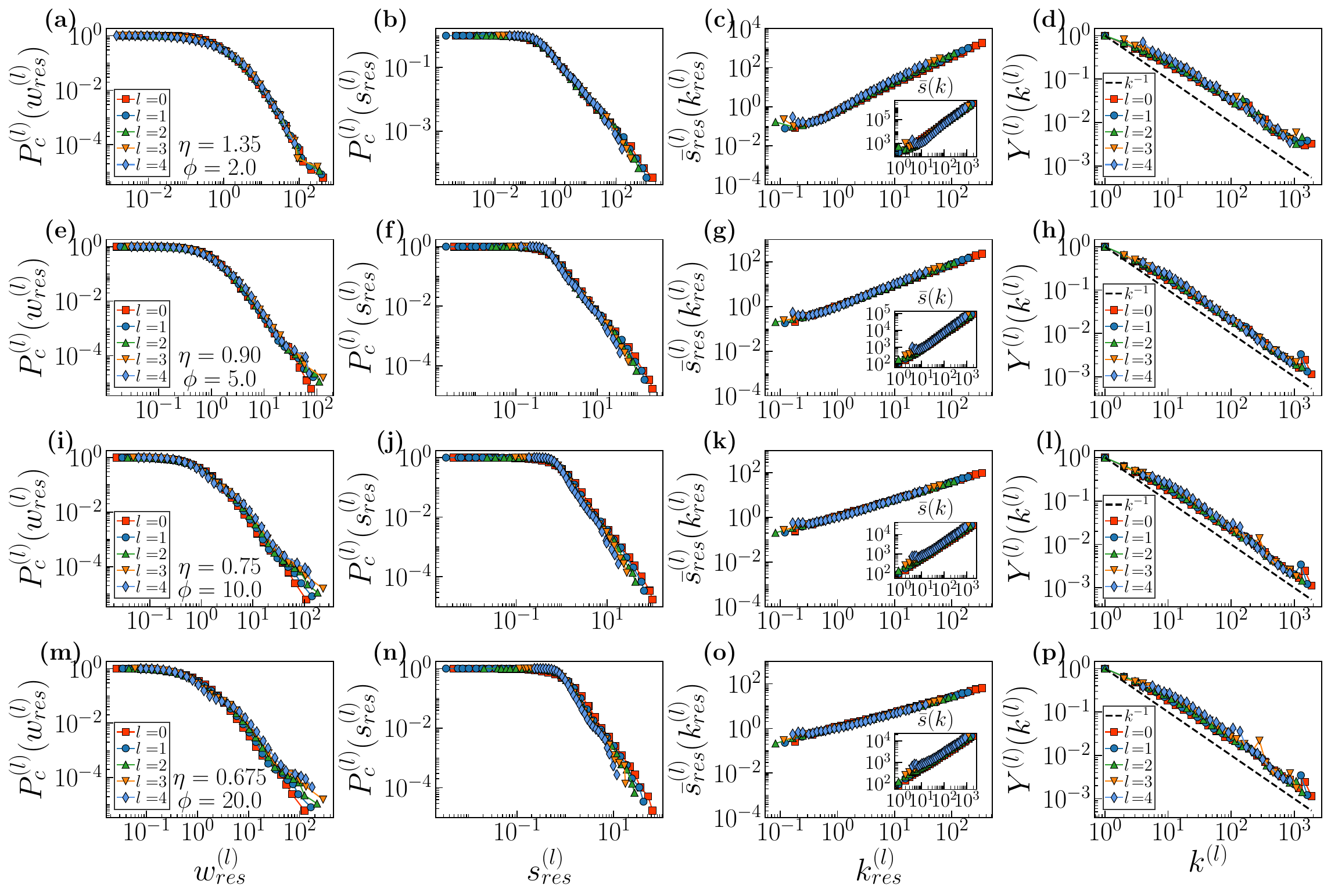}
	\caption{\textbf{$\phi$-GRW in synthetic network with different  $\eta$.} First column: complementary cumulative weight distributions $P_c^{(l)}(w_{res}^{(l)})$ of rescaled weights $w^{(l)}_{res}=w^{(l)}/\langle w^{(l)}\rangle$ for different layers $l$.
		Second column: complementary cumulative strength distributions $P_c^{(l)}(s_{res}^{(l)})$ of rescaled strengths $s^{(l)}_{res}=s^{(l)}/\langle s^{(l)}\rangle$ for different layers $l$.
		Third column: average rescaled strengths as a function of rescaled degrees, i.e.,  $\bar{s}_{res}^{(l)}(k_{res}^{(l)})=\bar{s}^{(l)}(k)/\langle s\rangle^{(l)}$. Inset shows average strength as a function of degree.
		Last column: disparity of nodes as a function of their degree. Each row indicates a synthetic network.The parameters are $N=58948$, $\beta= 1.5$, $\mu=0.0413$, $a=100$, $\eta=1.0$, $\gamma=2.5$ and $\langle \epsilon^2 \rangle=1.0$.}  
\end{figure}

\clearpage
\newpage
\section{The asymptotic behavior of the $\phi$-norm}
To better understand the behavior of $\phi$ normalization in different weighted distributions, we perform the following test. We generate the sampled weights $\omega_{mn}$ according to the weight distribution $p(\omega_{mn})\sim \omega_{mn} ^{-\delta}$. The smaller values of the exponent $\delta$, the more heterogeneous of the distributions.With the sampled weights $\omega_{mn}$ on hand, we can normalize the weight as: 
\begin{equation} 
	\label{omega_pri}
	\omega'(\phi)= \left[ \sum\limits_{e=1}^{E} \left( \omega_{mn} \right)^{\phi} \right]^{1/\phi},
\end{equation}
where $E$ is the number of sampled weights combining into new weight $\omega'$.  
The implementation process is as follows.

(1) We firstly generate a weight list $w_{list}$ with sampled length $N=20000$ from distribution $p(\omega_{mn})\sim \omega_{mn} ^{-\delta}$.

(2) We divide the weight list $w_{list}$ into non-overlapping groups in sequence, where each group's size equals $E$. In other words, each group has $E$ samples $\omega_{mn}$. We then calculate $\omega'(\phi)$ wit Eq.(\ref{omega_pri}) in each group for different $\phi$.

(3) We compare $\omega'(\phi)$ with $\omega'(\phi=1)$ and $\omega'(\phi=\infty)$. Note that sum-GRW corresponds to the case $\phi=1$ while sup-GRW to $\phi=\infty$.

We have two ways to check the asymptotic behavior of the $\phi$-norm in the empirical weighted distributions. The first one is as simple as the one in synthetic distribution. We only need to replace the weight list $w_{list}$ with the empirical data. In this case, the samples $\omega_{mn}$ in each group are uncorrelated. However, in the $\phi$-GRW process, the weights in the same group may relate to the coordinates of sub-nodes $m$ and $n$. Therefore, we implement the second way to check the asymptotic behavior. 

(1) We implement the $\phi$-GRW process with one layer. There may have $E$ weights links between the constituent nodes of two supernodes. Note that when $r=2$, the number of links $E$ could be $1$, $2$, $3$ or $4$. So, we divide the empirical weights into different group, where each group's size equals $E$.

(2) We then calculate $\omega'(\phi)$ wit Eq.(\ref{omega_pri}) in each group for different $\phi$.

(3) We compare $\omega'(\phi)$ with $\omega'(\phi=1)$ and $\omega'(\phi=\infty)$. Note that sum-GRW corresponds to the case $\phi=1$ while sup-GRW to $\phi=\infty$.

In the end, we find that the results obtained by these two ways are robust. We only show the results for the sets following the coarse-graining (i.e., the second way) in this paper.
\clearpage
\newpage
\subsubsection{Synthetic weighted distributions}
\begin{figure}[ht] 	
	\centering
	\includegraphics[width=0.6\linewidth]{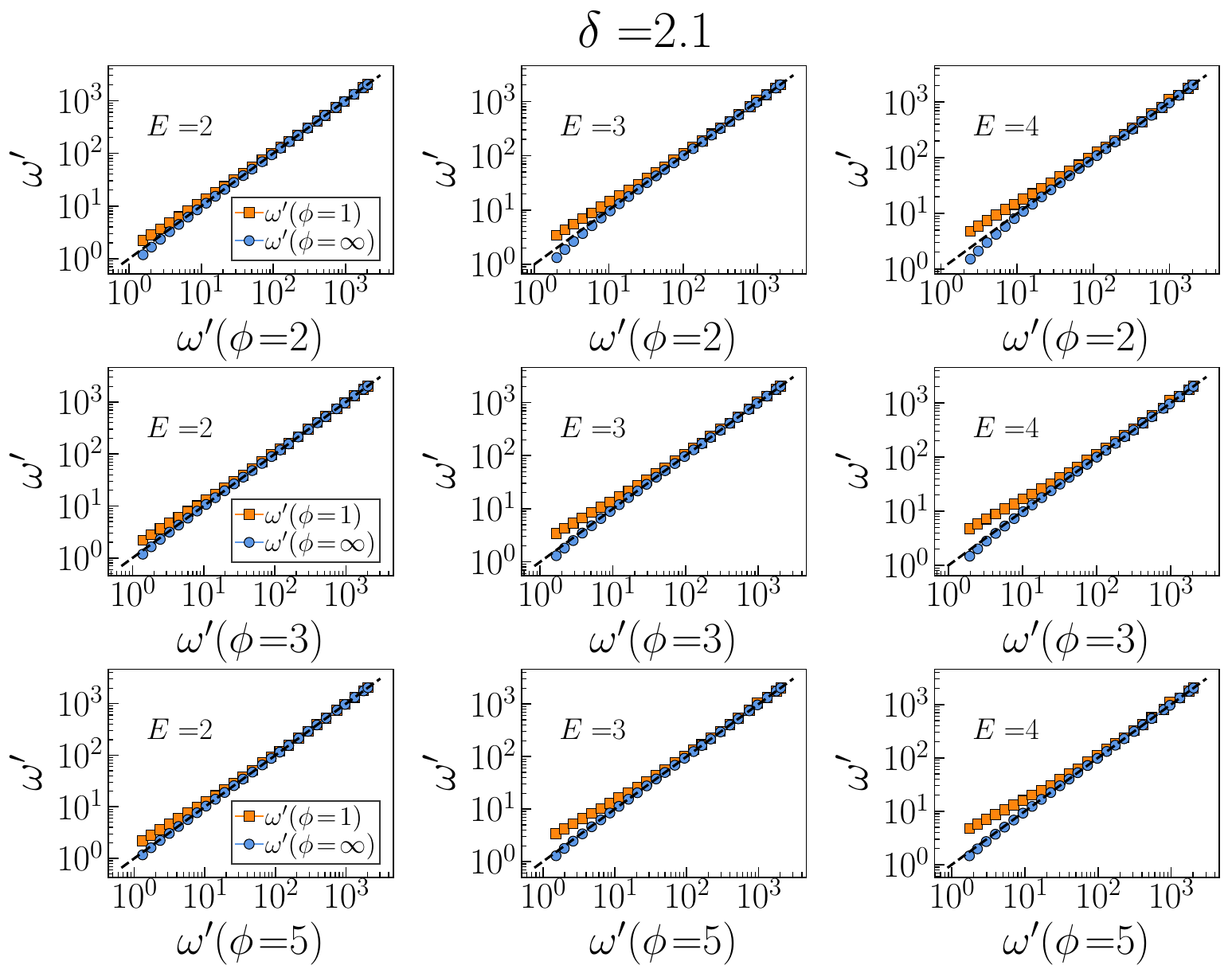} 
	\caption{The normalized weight $\omega'(\phi=1)$ and $\omega'(\phi=\infty)$ versus $\omega'(\phi)$ for different $E$ and $\phi$. Note that sum-GRW corresponds to the case $\phi=1$ while sup-GRW to $\phi=\infty$.}
\end{figure}

\begin{figure}[ht] 	
	\centering
	\includegraphics[width=0.6\linewidth]{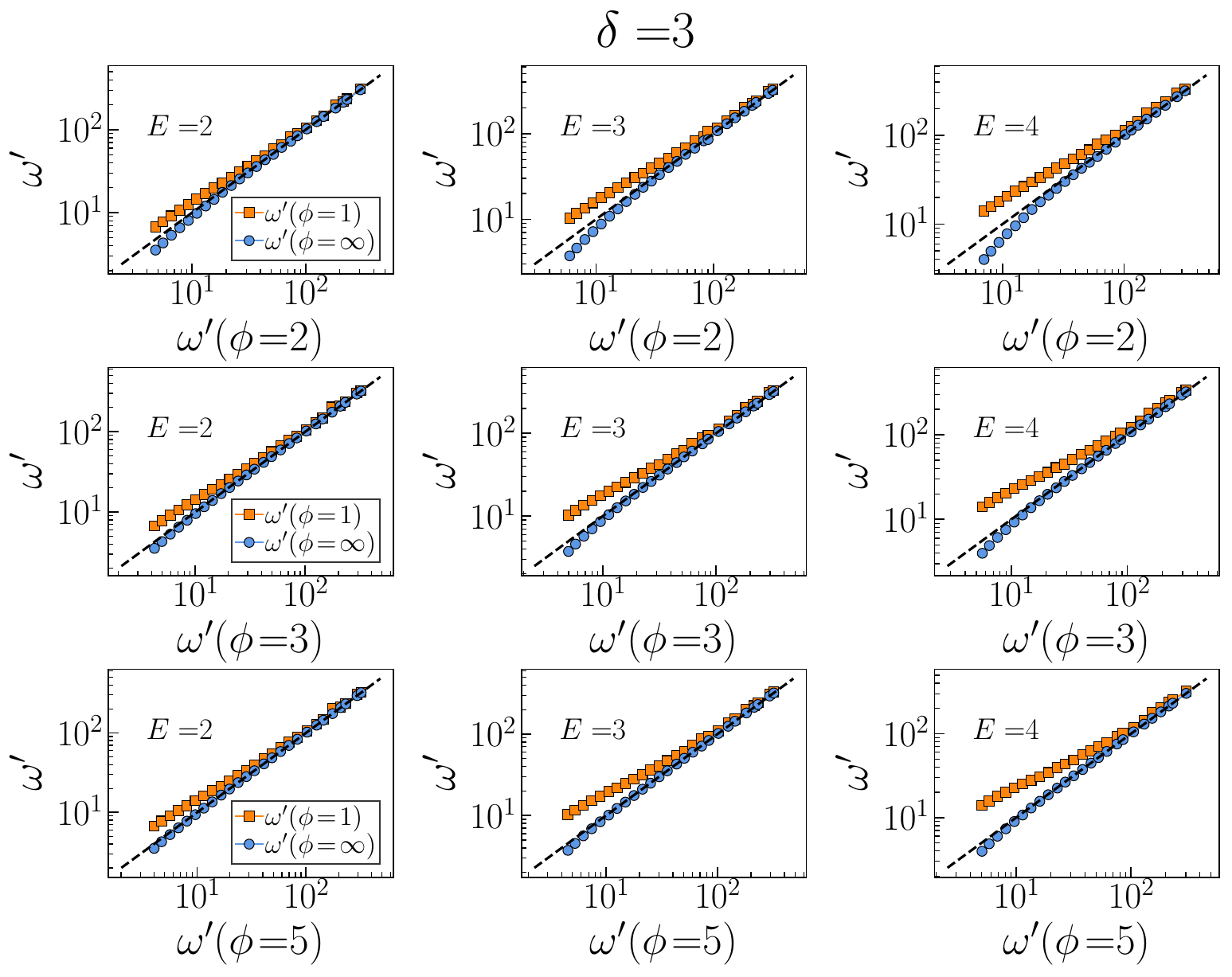} 
	\caption{The normalized weight $\omega'(\phi=1)$ and $\omega'(\phi=\infty)$ versus $\omega'(\phi)$ for different $E$ and $\phi$. Note that sum-GRW corresponds to the case $\phi=1$ while sup-GRW to $\phi=\infty$.}
\end{figure}

\clearpage
\newpage
\subsubsection{Empirical weighted distributions}

\begin{figure}[ht] 	
	\centering
	\includegraphics[width=0.9\linewidth]{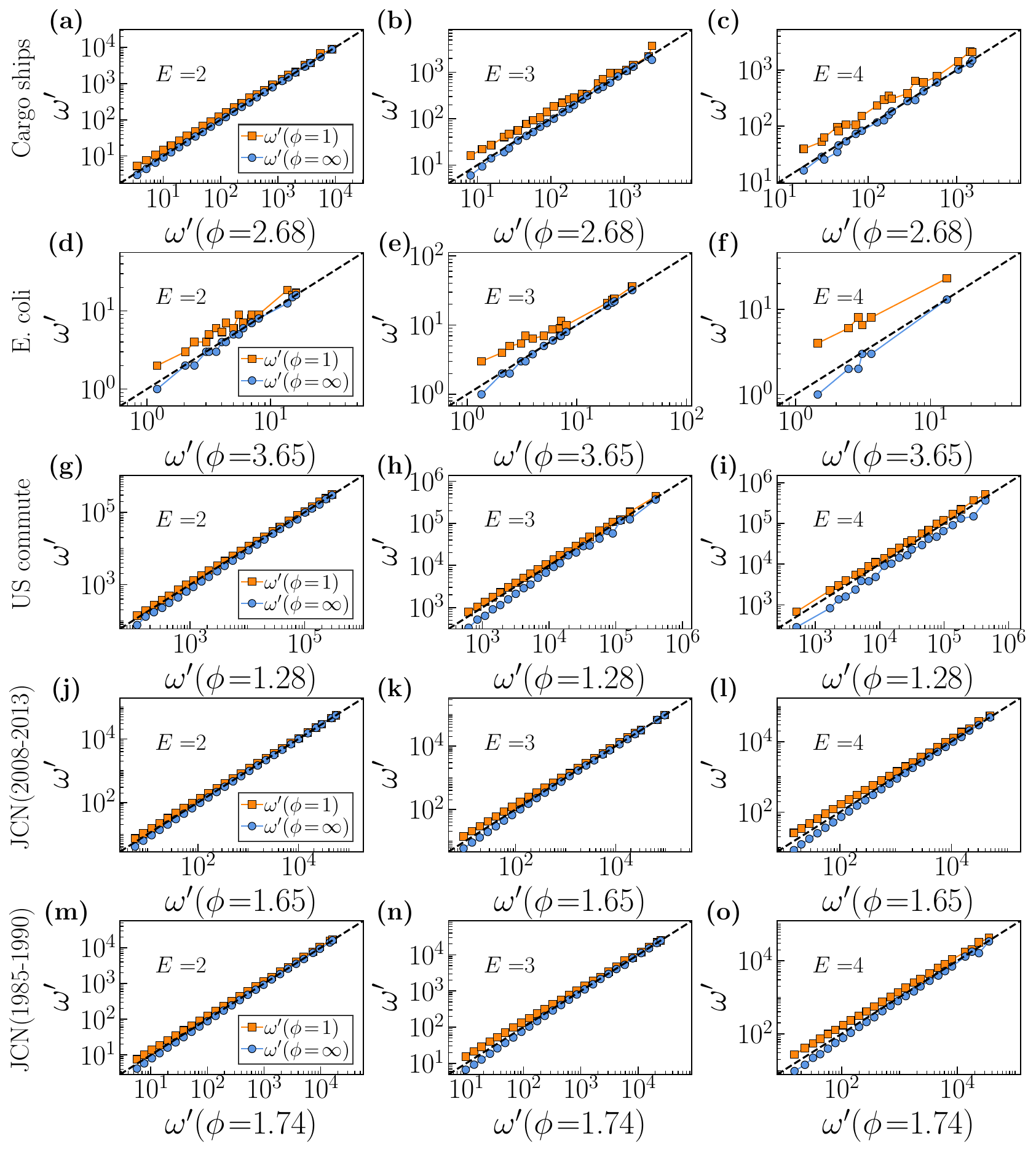} 
	\caption{The normalized weight $\omega'(\phi=1)$ and $\omega'(\phi=\infty)$ versus $\omega'(\phi)$ for different $E$ and inferred $\phi$. Note that sum-GRW corresponds to the case $\phi=1$ while sup-GRW to $\phi=\infty$.}
\end{figure}
\begin{figure}[ht] 	
	\centering	
	\includegraphics[width=1\linewidth]{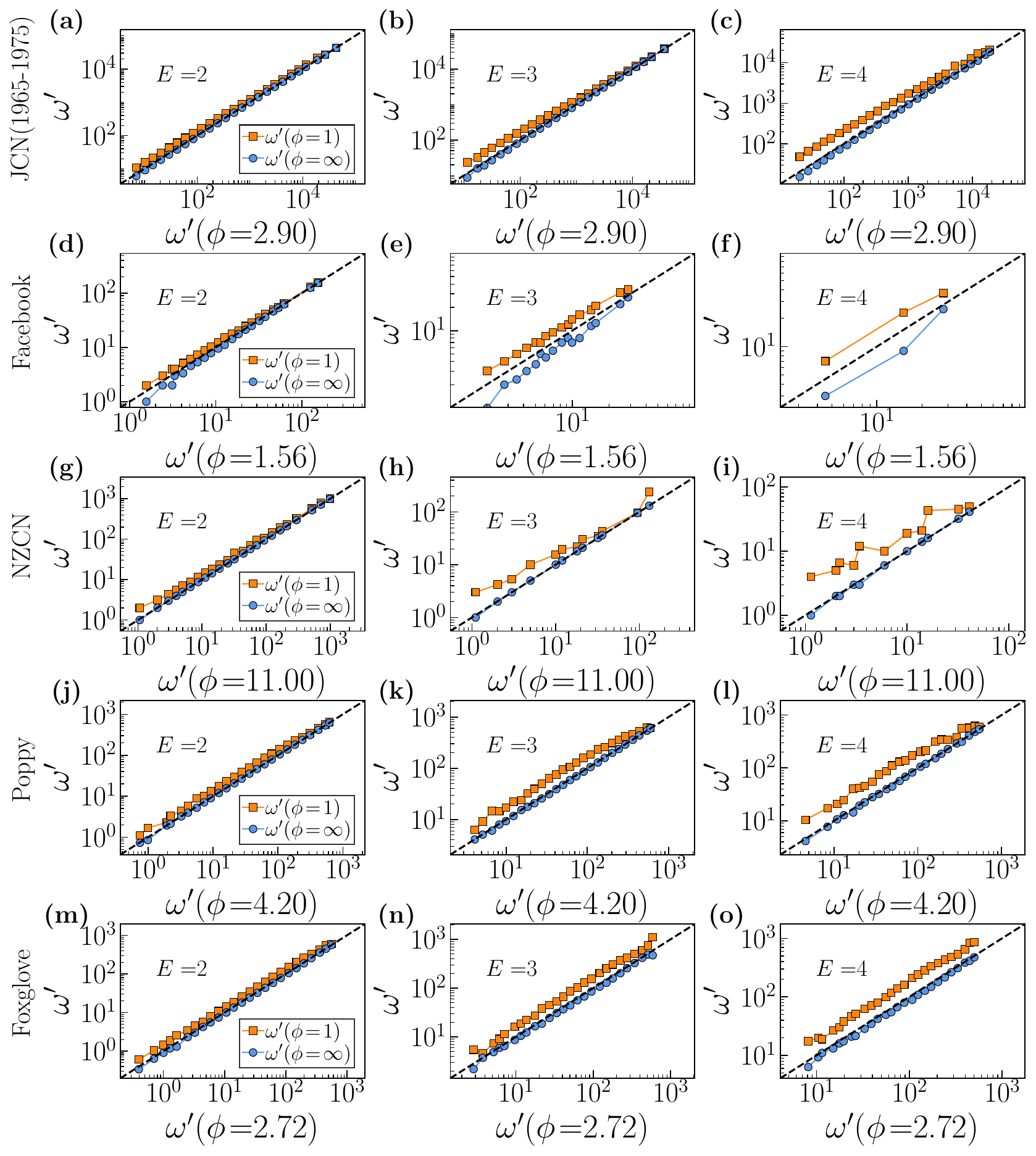} 
	\caption{The normalized weight $\omega'(\phi=1)$ and $\omega'(\phi=\infty)$ versus $\omega'(\phi)$ for different $E$ and inferred $\phi$. Note that sum-GRW corresponds to the case $\phi=1$ while sup-GRW to $\phi=\infty$.}
\end{figure}

\newpage
\clearpage
\section{Results for random-GRW }
A prescription selecting the weight between two supernodes at random from the coarse-grainable set would always result in the self-similarity of the distribution of weights if the selection set is supplemented with the links between nodes in the same supernode.
However, those links are coarse-grained in the renormalization process and the balance between the weights of links inside supernodes and of links between nodes in different supernodes dictates in which situations the random selection works. Experiments in synthetic networks, Figs.~S22 and S23, prove that the heterogeneity of the distribution of weights favors a better self-similar scaling. 
Decreasing the coupling of weights with topology and geometry in the W$\mathbb{S}^D$ model produces more homogeneous distributions of weights, 
which causes the loss of self-similarity in the flow, see SI Fig.~S24 for results relative to the metabolic network of {\it E. coli}. 

\subsection{Random-GRW in synthetic network }
\begin{figure}[ht] 
	\includegraphics[width=1\linewidth]{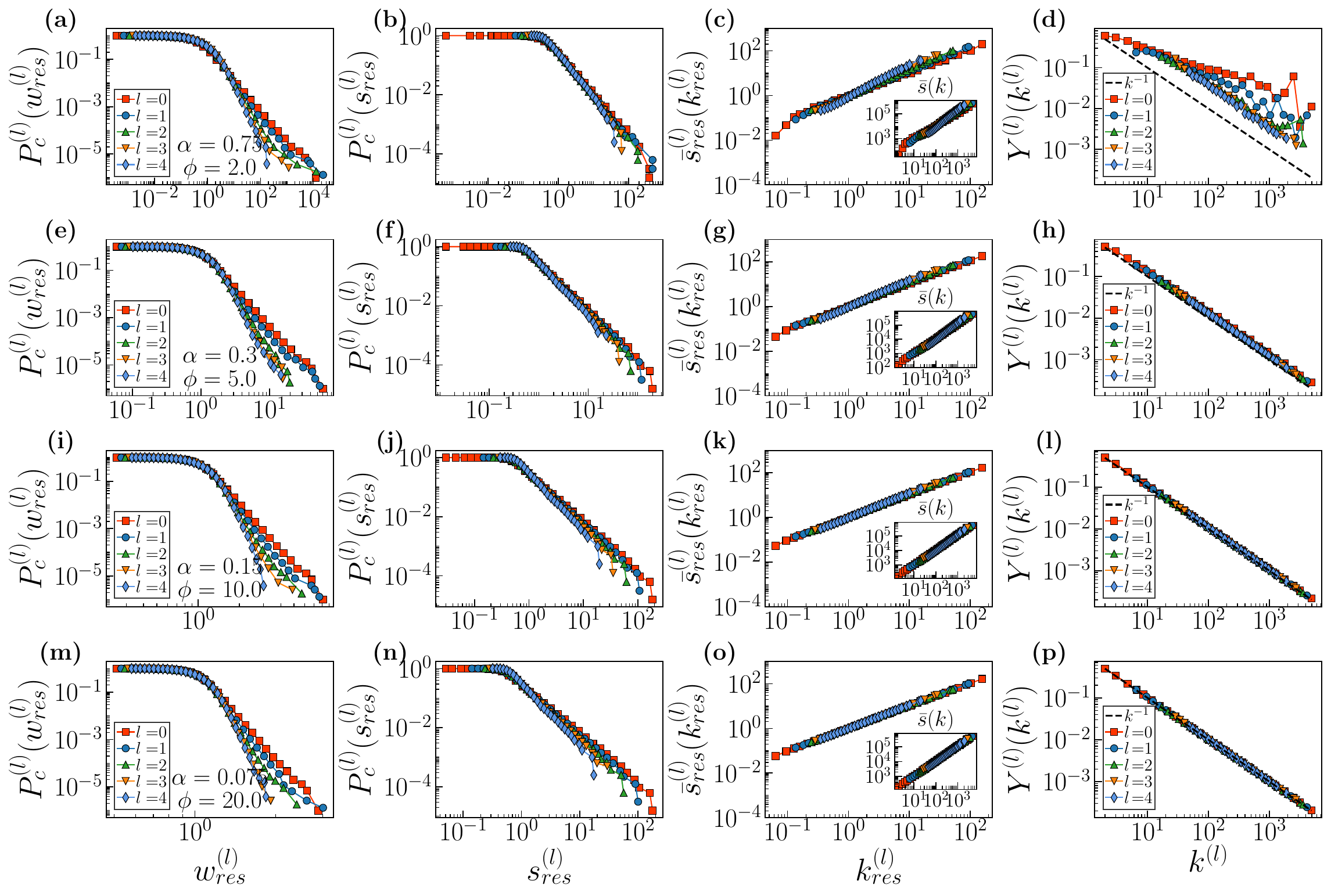}
	\caption{\textbf{Random-GRW in synthetic network with different $\alpha$.} First column: complementary cumulative weight distributions $P_c^{(l)}(w_{res}^{(l)})$ of rescaled weights $w^{(l)}_{res}=w^{(l)}/\langle w^{(l)}\rangle$ for different layers $l$.
		Second column: complementary cumulative strength distributions $P_c^{(l)}(s_{res}^{(l)})$ of rescaled strengths $s^{(l)}_{res}=s^{(l)}/\langle s^{(l)}\rangle$ for different layers $l$.
		Third column: average rescaled strengths as a function of rescaled degrees, i.e.,  $\bar{s}_{res}^{(l)}(k_{res}^{(l)})=\bar{s}^{(l)}(k)/\langle s\rangle^{(l)}$. Inset shows average strength as a function of degree.
		Last column: disparity of nodes as a function of their degree. Each row indicates a synthetic network.The parameters are The parameters are $N=64000$,  $\beta= 1.5$, $\mu=0.0059$, $a=100$, $\eta=1.0$, $\gamma=2.5$, $\langle k \rangle=35$ and $\langle \epsilon^2 \rangle=1.0$.}  
\end{figure}

\begin{figure}[ht] 
	\includegraphics[width=1\linewidth]{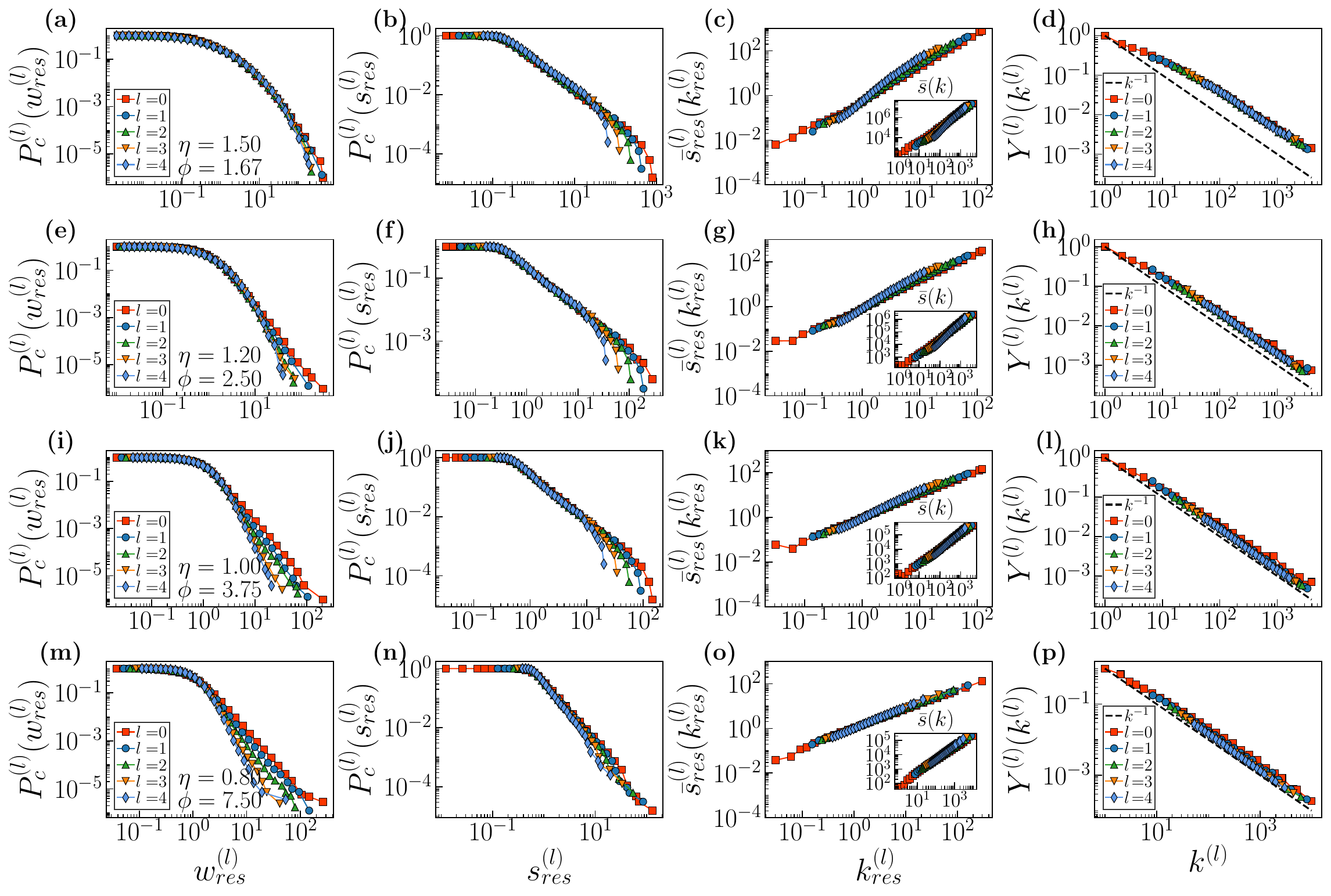}
	\caption{\textbf{Random-GRW in synthetic network with different $\eta$.} First column: complementary cumulative weight distributions $P_c^{(l)}(w_{res}^{(l)})$ of rescaled weights $w^{(l)}_{res}=w^{(l)}/\langle w^{(l)}\rangle$ for different layers $l$.
		Second column: complementary cumulative strength distributions $P_c^{(l)}(s_{res}^{(l)})$ of rescaled strengths $s^{(l)}_{res}=s^{(l)}/\langle s^{(l)}\rangle$ for different layers $l$.
		Third column: average rescaled strengths as a function of rescaled degrees, i.e.,  $\bar{s}_{res}^{(l)}(k_{res}^{(l)})=\bar{s}^{(l)}(k)/\langle s\rangle^{(l)}$. Inset shows average strength as a function of degree.
		Last column: disparity of nodes as a function of their degree. Each row indicates a synthetic network.The parameters are The parameters are $N=64000$,  $\beta= 1.5$, $\mu=0.0059$, $a=100$, $\alpha=0.40$, $\gamma=2.5$, $\langle k \rangle=35$ and $\langle \epsilon^2 \rangle=1.0$.}  
\end{figure}
\clearpage
\newpage
\subsection{Random-GRW in E. coli network}
\begin{figure}[!h] 	
	\includegraphics[width=1\linewidth]{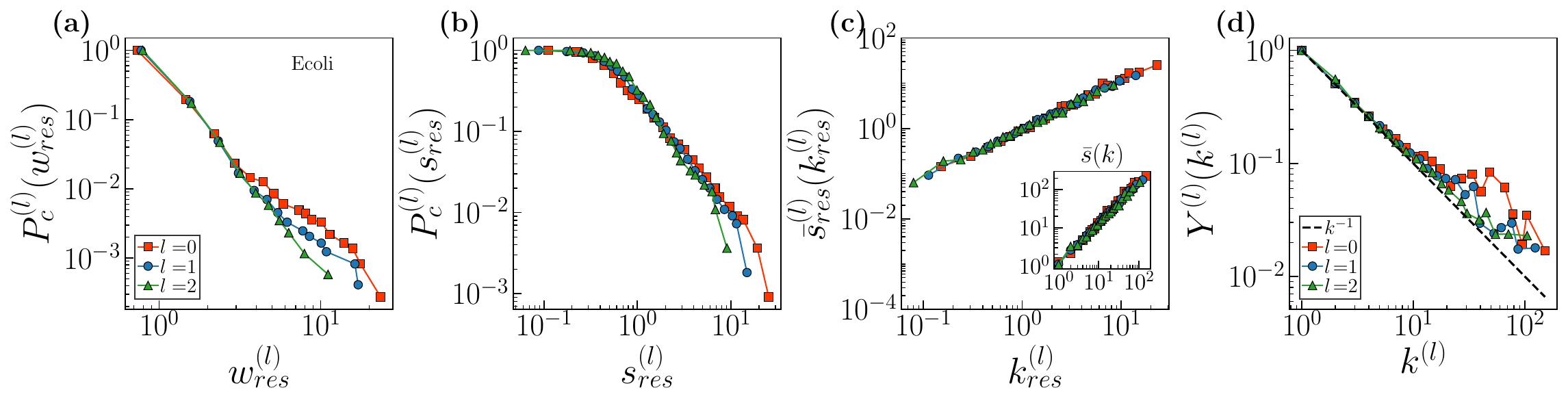} 
	\caption{\textbf{random-GRW in E. coli network.}  
		(a) Complementary cumulative weight distributions $P_c^{(l)}(w_{res}^{(l)})$ of rescaled weights $w^{(l)}_{res}=w^{(l)}/\langle w^{(l)}\rangle$ for different layers $l$. (b) Complementary cumulative strength distributions $P_c^{(l)}(s_{res}^{(l)})$ of rescaled strengths $s^{(l)}_{res}=s^{(l)}/\langle s^{(l)}\rangle$ for different layers $l$. (c) average rescaled strengths as a function of rescaled degrees, i.e.,  $\bar{s}_{res}^{(l)}(k_{res}^{(l)})=\bar{s}^{(l)}(k)/\langle s\rangle^{(l)}$. Inset shows average strength as a function of degree. (d) Disparity of nodes as a function of their degree. }
\end{figure}
\newpage
\clearpage
\section{Scaled down replicas of weighted networks with sup-GRW}
\begin{figure}[ht]
	\includegraphics[width=0.8\linewidth]{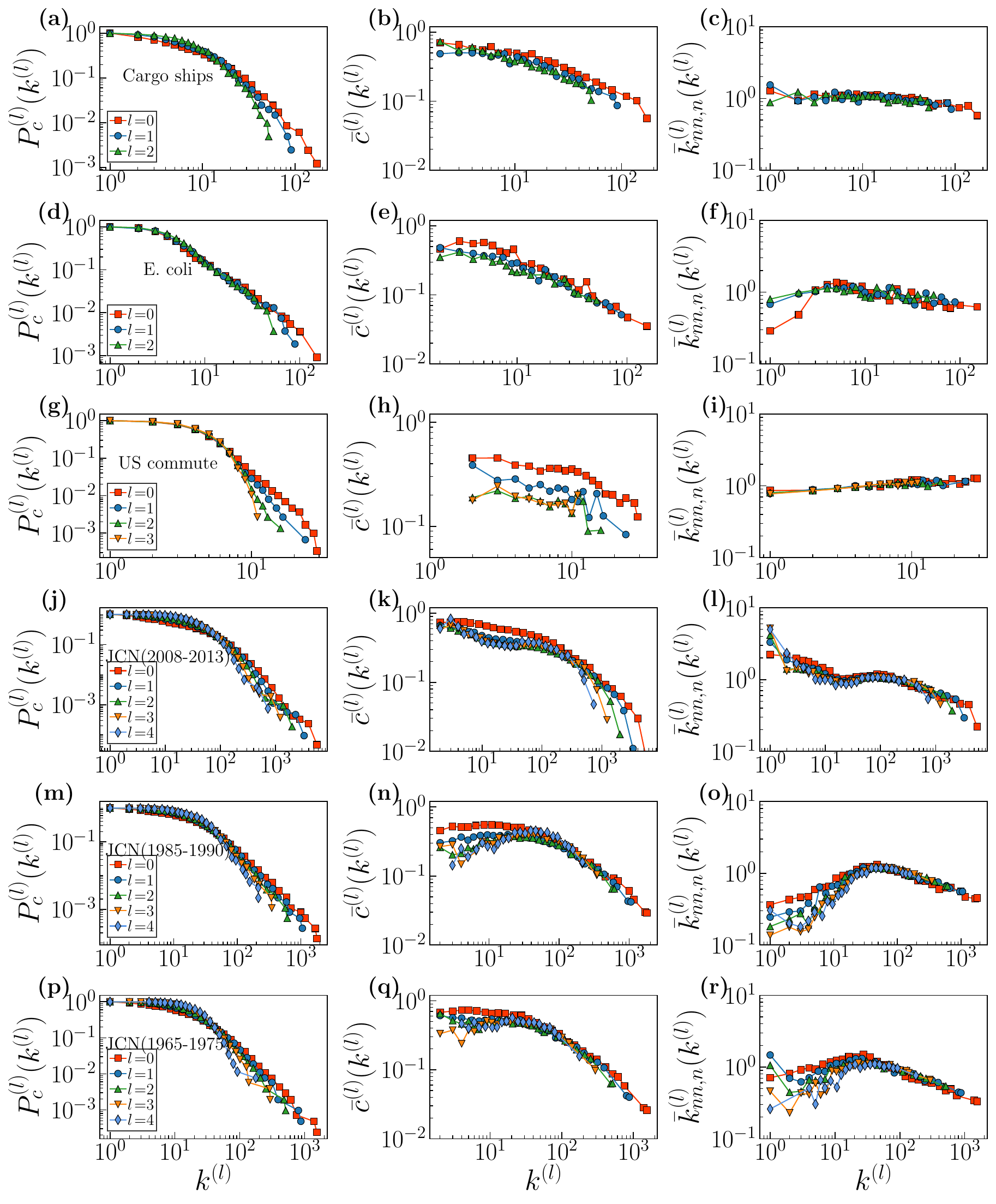}
	\caption{\textbf{Scaled down replicas with sup-GRW in different empirical networks.} First column: complementary cumulative degree distribution $P_c^{(l)}(k^{(l)})$, Second column: the degree-dependent clustering coefficient $\bar{c}^{(l)}(k^{(l)})$, Last column: normalized average nearest-neighbour degree $\bar{k}_{nn,n}^{(l)} (k^{(l)}) = \bar{k}_{nn}^{(l)} (k^{(l)}) \langle  k^{(l)}\rangle/\langle(k^{(l)})^2\rangle$ of degrees $k^{(l)}$ for different layers $l$. Each row indicates an empirical network.}
\end{figure}

\begin{figure}[ht] 	
	\includegraphics[width=0.8\linewidth]{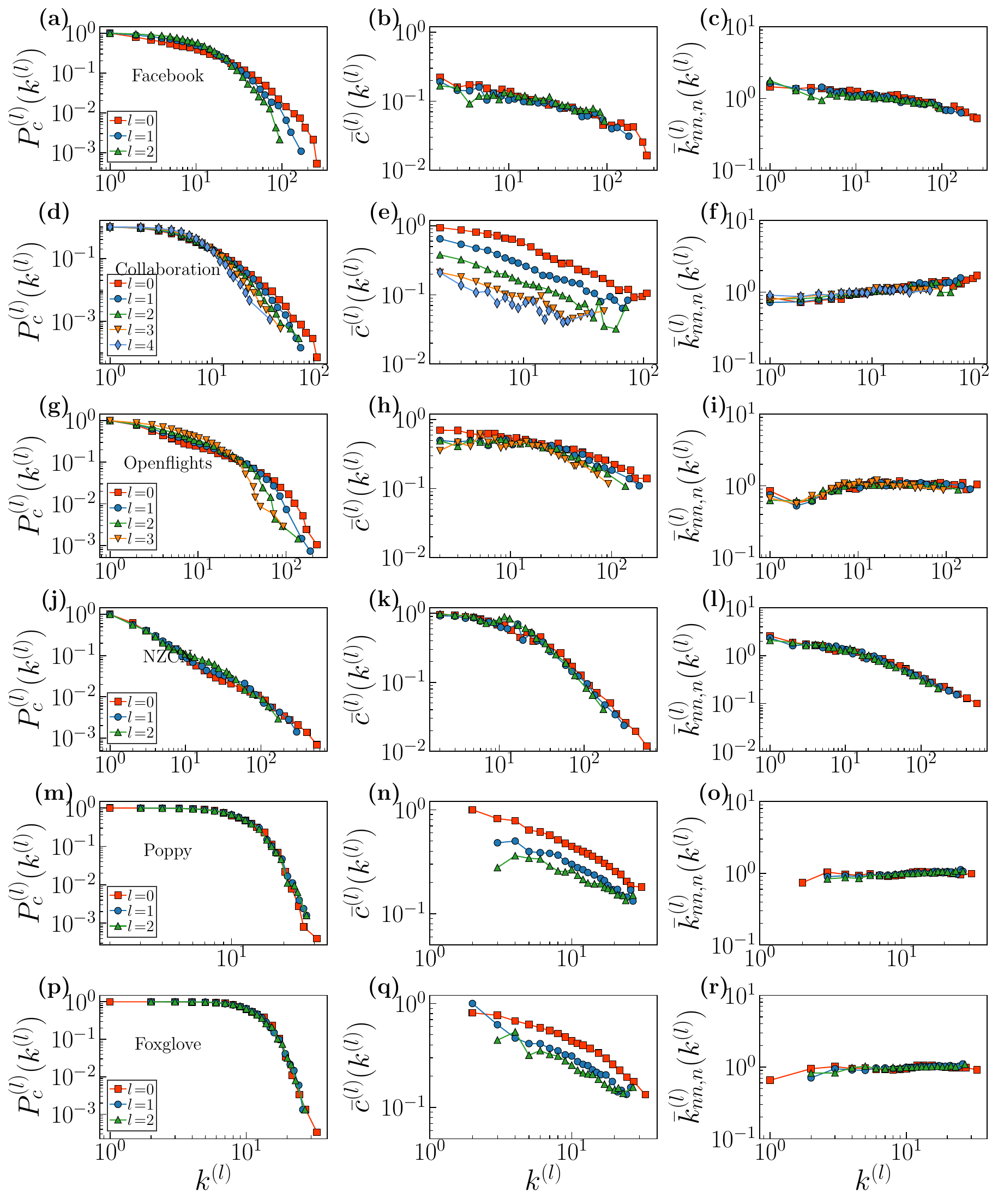} 
	\caption{\textbf{Scaled down replicas with sup-GRW in different empirical networks.} First column: complementary cumulative degree distribution $P_c^{(l)}(k^{(l)})$, Second column: the degree-dependent clustering coefficient $\bar{c}^{(l)}(k^{(l)})$, Last column: normalized average nearest-neighbour degree $\bar{k}_{nn,n}^{(l)} (k^{(l)}) = \bar{k}_{nn}^{(l)} (k^{(l)}) \langle  k^{(l)}\rangle/\langle(k^{(l)})^2\rangle$ of degrees $k^{(l)}$ for different layers $l$. Each row indicates an empirical network.}
\end{figure}

\begin{figure}[ht] 	
	\includegraphics[width=1\linewidth]{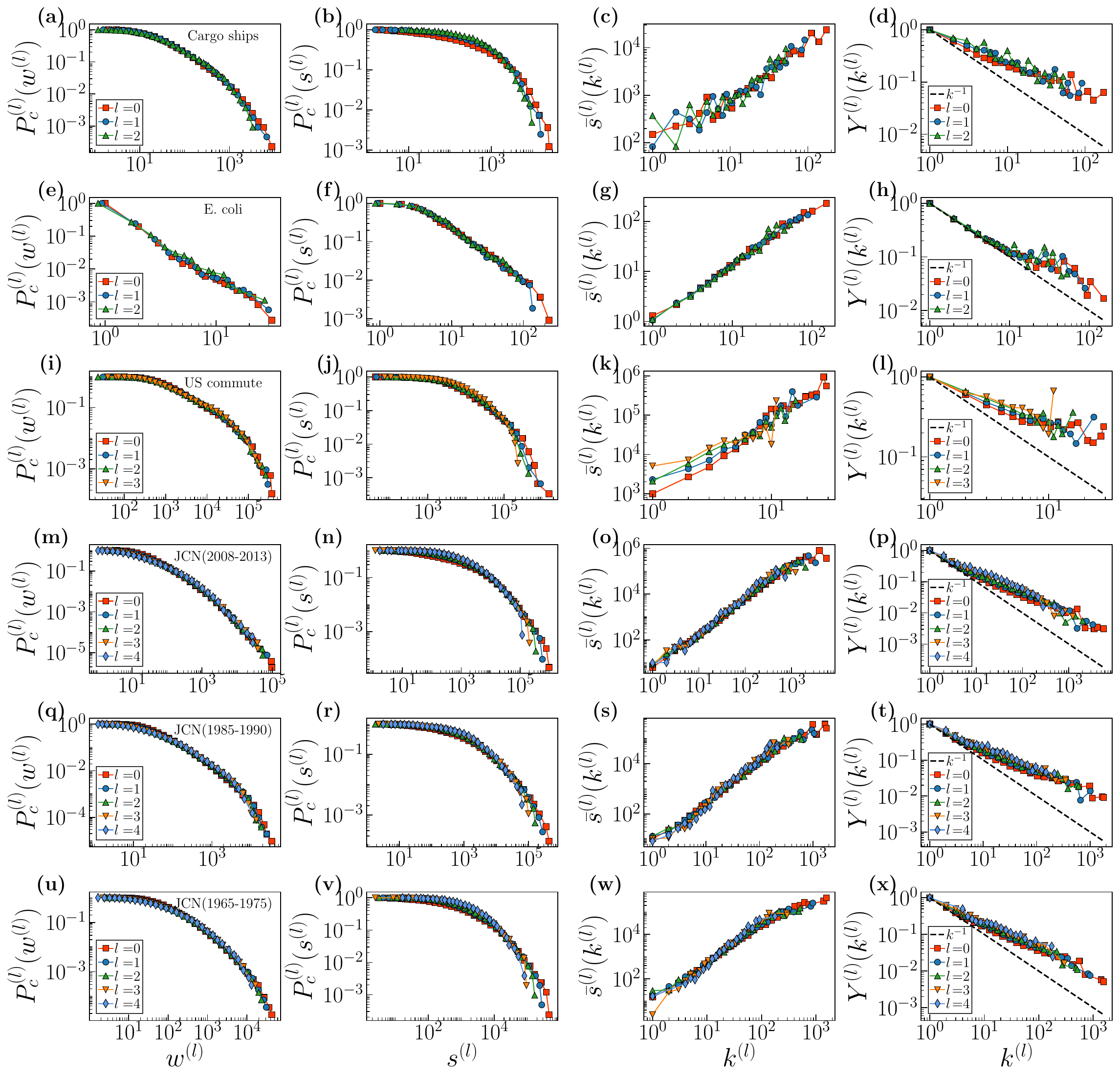} 
	\caption{\textbf{Scaled down replicas with sup-GRW in different empirical networks.} First column: complementary cumulative weight distributions $P_c^{(l)}(w^{(l)})$ for different layers $l$.
		Second column: complementary cumulative strength distributions $P_c^{(l)}(s^{(l)})$ for different layers $l$.
		Third column: average strength $\bar{s}^{(l)}(k^{(l)})$ as a function of degree $k^{(l)}$.
		Last column: disparity of nodes as a function of their degree. Each row indicates an empirical network.}
\end{figure}

\begin{figure}[ht] 	
	\includegraphics[width=1\linewidth]{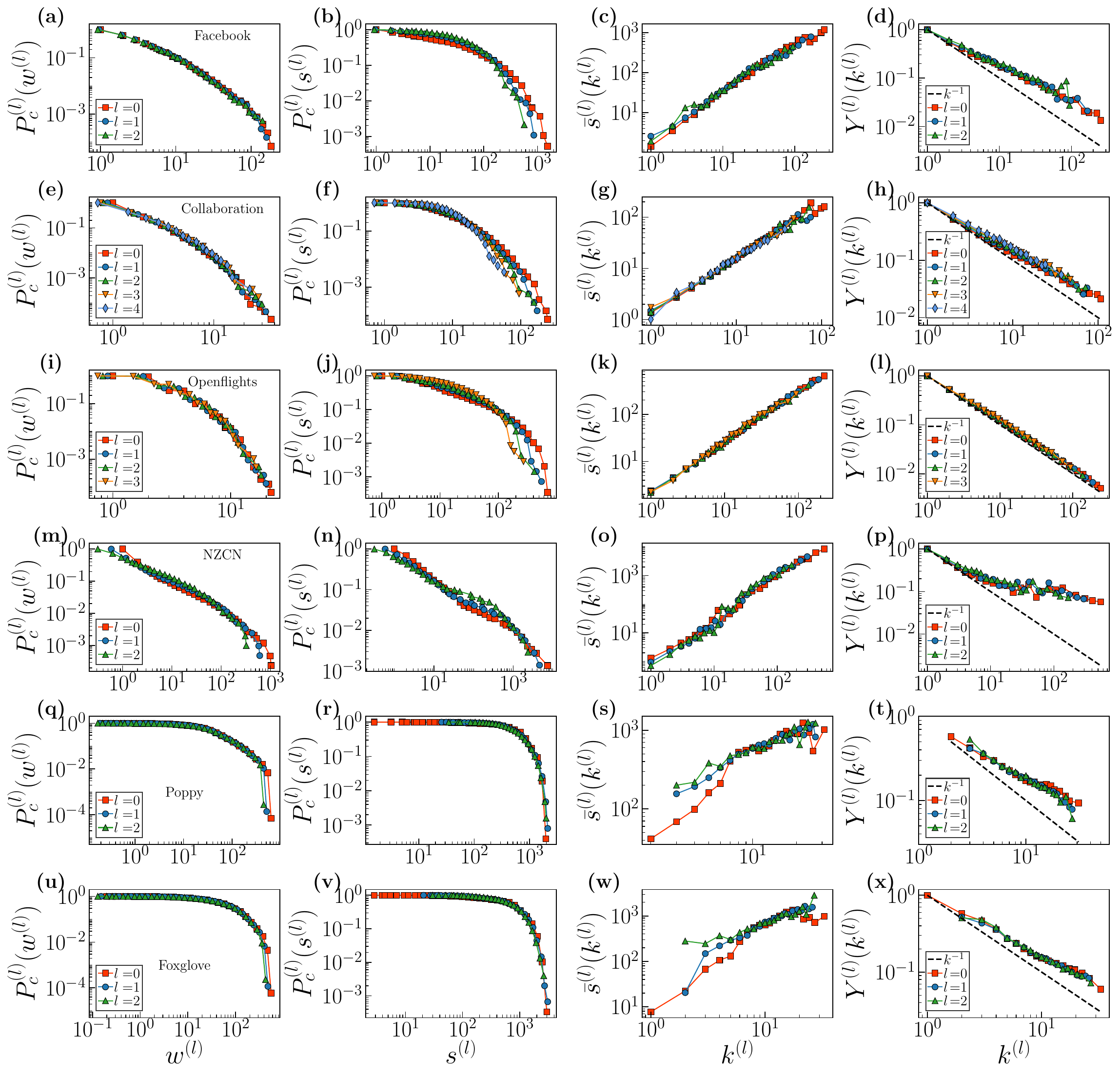} 
	\caption{\textbf{Scaled down replicas with sup-GRW in different empirical networks.} First column: complementary cumulative weight distributions $P_c^{(l)}(w^{(l)})$ for different layers $l$.
		Second column: complementary cumulative strength distributions $P_c^{(l)}(s^{(l)})$ for different layers $l$.
		Third column: average strength $\bar{s}^{(l)}(k^{(l)})$ as a function of degree $k^{(l)}$.
		Last column: disparity of nodes as a function of their degree. Each row indicates an empirical network.}
\end{figure}

\newpage
\clearpage
%

\end{document}